\documentclass[11pt]{article}

\usepackage{jcapmod}
\usepackage{graphicx}
\usepackage{dcolumn}
\usepackage{bm}
\usepackage{latexsym}
\usepackage{amsfonts}
\usepackage{cancel}
\usepackage{amssymb}
\usepackage{amsmath}
\usepackage{mathrsfs}
\usepackage{afterpage}
\usepackage{placeins}
\usepackage{color}
\usepackage{hyperref}
\usepackage{ulem}
\usepackage{natbib} 
\usepackage{siunitx}
\usepackage{verbatim}
\usepackage{ dsfont }
\def\be{\begin{equation}}
\def\ee{\end{equation}}
\def\bea{\begin{eqnarray}}
\def\eea{\end{eqnarray}}
\newcommand{\vs}{\nonumber\\}
\def\ba#1\ea{\begin{align}#1\end{align}}

\newcommand{\g}{\gamma}

\renewcommand{\v}[1]{\mathbf{#1}}
\newcommand{\vx}{\v{x}}
\newcommand{\vk}{\v{k}}
\newcommand{\vq}{\v{q}}
\newcommand{\bq}{\mathbf{q}}
\newcommand{\bk}{\mathbf{k}}
\newcommand{\bx}{\mathbf{x}}

\newcommand{\bn}{\mathbf{n}}

\newcommand{\vnhat}{\v{\hat{n}}}

\def\P{\mathcal{P}}

\def\Del{\eth}

\def\vnhat{\hat{\v{n}}}

\definecolor{green2}{cmyk}{1, 0, 1, 0.1}

\begin{document}

\begin{titlepage}
	\baselineskip=15.5pt \thispagestyle{empty}
	
	\bigskip\
	
	\vspace{1cm}
	\begin{center}
		{\fontsize{15.5}{10}\selectfont \sffamily \bfseries Primordial Gravitational Waves from Galaxy Intrinsic Alignments}	%\\[8pt]	%\\[12pt]
		% Preliminary title...
		% Inflationary Features in the BAO Spectrum
		% Inflationary Oscillations in the BAO Spectrum
		% Oscillatory Features in the BAO Spectrum
		% Features of the Early Universe in the BAO Spectrum
		% Featuring the Early Universe in the BAO Spectrum
		% Featuring the Primordial Universe in the BAO Spectrum
	\end{center}

	%\vspace{0.2cm}
	\begin{center}
		%{\fontsize{12}{30}\selectfont Daniel Baumann,$^{1}$ Florian Beutler,$^{2,3}$ Matteo Biagetti,$^{1}$\\[4pt]Daniel Green,$^{4}$ An\v{z}e Slosar$^{5}$ and Benjamin Wallisch$^{6,7,1,4}$}
		{\fontsize{12}{30}\selectfont Matteo Biagetti,$^{1}$ Giorgio Orlando,$^{1,2,3}$} 
	\end{center}
	
	\begin{center}
		%\vskip8pt
		%\textsl{$^1$ Institute of Theoretical Physics, University of Amsterdam, Amsterdam, The Netherlands}		%Institute for Theoretical Physics?
		
		\vskip8pt
		%\textsl{$^2$ Institute of Cosmology \& Gravitation, University of Portsmouth, Portsmouth, PO1 3FX, UK}
		
		\vskip8pt
		\textsl{ $^1$ Institute for Theoretical Physics, University of Amsterdam, 1098 XH Amsterdam, NL \\
	    $^2$ Dipartimento di Fisica e Astronomia “G. Galilei”, Universit\`{a} di Padova, I-35131, Padova, IT \\
		$^3$ INFN, Sezione di Padova, I-35131, Padova, IT}
			\vskip8pt
	\end{center}

	\vspace{1.2cm}
	\hrule \vspace{0.3cm}
	\noindent {\sffamily \bfseries Abstract}\\[0.1cm]
	Galaxy shapes have been observed to align with external tidal fields generated by the large-scale structures of the Universe. While the main source for these tidal fields is provided by long-wavelength density perturbations, tensor perturbations also contribute with a non-vanishing amplitude at linear order. We show that parity-breaking gravitational waves produced during inflation leave a distinctive imprint in the galaxy shape power spectrum which is not hampered by any scalar-induced tidal field. We also show that a certain class of tensor non-Gaussianities produced during inflation can leave a signature in the density-weighted galaxy shape power spectrum. We estimate the possibility of observing such imprints in future galaxy surveys.
	\vskip10pt
	\hrule
	\vskip10pt
\end{titlepage}

\thispagestyle{empty}
\setcounter{page}{2}
\tableofcontents
% All sections etc.\ are preliminary and only some rough ideas!

\clearpage

\section{Introduction}

The statistical distribution of galaxy shapes in the sky provides a great deal of astrophysical and cosmological information and has been used as a major observational probe in weak lensing studies \cite{Kaiser:1991qi,Bartelmann:1999yn,Peacock:2006kj,Munshi:2006fn,Hoekstra:2008db,Weinberg:2012es}. The image of gravitationally lensed galaxies is distorted near a foreground mass and the statistical study of these distortions allows to map the distribution of matter in the Universe in an unbiased way, therefore providing an important complementary probe to biased observations, such as galaxy number counts. Over the last decades, it has been realized that a major systematic in weak lensing measurements is introduced if galaxy shapes are intrinsically correlated \cite{Brainerd:1995da}. Pioneering work in trying to model these correlations was done in the early 2000s \cite{Catelan:2000vm,Croft:2000gz,Heavens:2000ad,Crittenden:2000wi,Hirata:2004gc} and refined later on \cite{Blazek:2011xq,Schmidt:2012nw,Schmidt:2013gwa,Joachimi:2013una,Chisari:2013dda,Blazek:2015lfa,Blazek:2015hqa,Schmidt:2015xka,Chisari:2016xki,Blazek:2017wbz,Schmitz:2018rfw,Okumura:2019ozd,Vlah:2019byq,Okumura:2019ned}. The existence of such intrinsic correlations is supported by their observation on luminous red galaxies at low redshift from the 2SLAQ and SDSS surveys \cite{Hirata:2007np, Okumura:2008du, Singh:2014kla,Singh:2015sva} and by recent measurements of the gravitational lensing-intrinsic alignment cross-correlation on BOSS survey data performed by \cite{Pedersen:2019wfp}. 

In the effort of modelling intrinsic alignments as a systematic effect, it has been realized that they themselves contain valuable cosmological information. Indeed, the intrinsic shape of a galaxy correlates with the large-scale structures of the Universe and therefore it traces the three-dimensional distribution of the matter density field on large scales. Similarly to other probes, information on how the matter density field is correlated over long distances is not only useful for cosmological parameter inference, but also can provide constraints on early universe physics. There are two primordial signatures that can leave an imprint in galaxy alignments in this way: first, inflationary bispectra of the primordial curvature perturbation, also known as primordial non-Gaussianity, contribute to the galaxy shape power spectrum in a similar way as for the ``scale-dependent bias'' in the case of galaxy clustering searches  (see \cite{Biagetti:2019bnp} for a recent review). The primordial bispectra which have been considered in the context of intrinsic aligments are the so-called local-type primordial non-Gaussianity  \cite{Chisari:2013dda} and subsequently, more broadly, models with a sizeable anisotropic squeezed limit from scalar and higher-spin fields \cite{Schmidt:2015xka,Chisari:2016xki,Kogai:2018nse}. Secondly, primordial gravitational waves source intrinsic alignments at leading order. This was argued early on in \cite{Dodelson:2003bv,Dodelson:2010qu} and later elaborated in a complete framework \cite{Schmidt:2012ne,Schmidt:2012nw,Schmidt:2013gwa,Chisari:2014xia}.
%In this case, correlations of the galaxy shape field projected on the sky are decomposed into E and B modes, similarly to what usually done in CMB observations.

While a number of observables at different cosmological stages are sensitive to signatures of primordial non-Gaussianity (see \cite{Meerburg:2019qqi} for a recent overview of probes), the prospect for observing primordial gravitational waves in the future almost exclusively lies on CMB observations, such as the Simons Observatory \cite{Ade:2018sbj}, LITEBird \cite{Hazumi:2019lys} and CMB Stage 4 \cite{Abazajian:2016yjj}. Upcoming galaxy imaging surveys, such as Euclid \cite{Amendola:2016saw} and the Vera C. Rubin Observatory \cite{Ivezic:2008fe}, will provide an unprecedented dataset consisting of millions of galaxy shapes. It is therefore imperative to understand in detail how to exploit this wealth of information as an alternative probe of primordial gravitational waves using galaxy intrinsic alignments.  

The way in which tensor modes affect intrinsic galaxy shapes can be understood by observing that the leading locally observable effect of a long-wavelength perturbation $k_L$, be it a tensor or a scalar one, on a region of size much smaller than $1/k_L$ is an effective tidal field \cite{Schmidt:2013gwa}, causing the deformation of the galaxy shapes with respect to the rotational symmetry. In the case of tensors, the tidal force is generated by an effective peculiar potential of the form $\psi^F=-1/4(\ddot{h}_{ij}+2H \dot{h}_{ij})x^i_Fx^j_F$, where $h_{ij}$ are the transverse and traceless tensor perturbations around a Friedman, Lema\"itre, Robertson and Walker (FLRW) metric and $F$ indicates the Fermi Normal Coordinate (FNC) frame,  in which the metric is Minkowski along the central geodesic passing through the center of mass of a given region  of  the  Universe \cite{Fermi:1922,Manasse:1963zz}. The galaxy shape field is then assumed to linearly respond to changes in the tidal field generated by these tensor perturbations, such that the galaxy shape power spectrum, projected on the sky and properly decomposed in spherical harmonics, exhibits non-zero E and B modes on large scales.

In this paper, we elaborate on the imprint of primordial gravitational waves on galaxy intrinsic alignments by making three main points:
\begin{itemize}
    \item B modes of the galaxy shape power spectrum are not only intrinsically sourced by tensor perturbations, but also by scalar perturbations, through the curvature of the gravitational potential. This fact is already known since \cite{Hirata:2004gc}, but up to now it has been only calculated in the flat-sky approximation and therefore not valid on the largest scales. Because for primordial gravitational waves the largest scales are crucial, we provide a full-sky calculation. We find that the scalar-induced intrinsic alignments are typically larger by 2-3 orders of magnitude at best, i.e. on the largest scales, than the ones sourced by primordial gravitational waves, therefore providing a large contaminant to the primordial signature.
    \item Parity-violating physics taking place during inflation can induce chiral gravitational waves which in turn source an E-B correlation in the galaxy shapes power spectrum. These parity-breaking contributions are not generated by scalar perturbations and therefore any signature of this E-B correlation in the data would be a smoking gun for parity breaking processes of primordial origin.
    \item Inflationary bispectra involving primordial tensor perturbations also source intrinsic alignments. We estimate which of these non-Gaussianities have a sizable impact on the galaxy shape power spectrum.
\end{itemize}
The calculation of the galaxy shape power spectra is performed by projecting the three-dimensional galaxy shape field on the sky and decomposing the two-dimensional quantity with spherical harmonics, using recently developed techniques \cite{Schmidt:2012ne,Schmidt:2012nw,Schmidt:2013gwa}.  We provide a full-sky computation of all quantities and we develop an approximate approach to highly oscillatory integrals which allows for fast computation of correlation functions of galaxy shapes at high $\ell$ and make our code public\footnote{\textrm{https://gitlab.com/mbiagetti/tensor\textunderscore fossil}}.

The structure of the paper is as follows: we review past and recent progress on galaxy intrinsic alignments in Section  \ref{sec:scalar}. We then explain in detail how to compute the effect of primordial tensor perturbations on galaxy shapes in Section  \ref{sec:tensors}, arguing that primordial B modes are challenging to be constrained using intrinsic alignments due to a contamination from scalar-induced alignments. We make the point in Section  \ref{sec:chiral} that parity breaking primordial gravitational waves are not affected by this contamination and provide a pristine window into primordial processes using the EB correlation of galaxy shapes on large scales. We finally argue that tensor non-Gaussianities can also source the galaxy shape power spectrum and provide estimation of this signature for two promising models in Section  \ref{Imprint_no_Gaus} and make final remarks in Section  \ref{sec:conclusions}.

\section{Overview of galaxy intrinsic alignments}\label{sec:scalar}

In this section, we review the general formalism required to compute correlations between intrinsic galaxy shapes and large-scale tidal fields generated by the gravitational potential, mostly summarizing known results from previous literature (see \cite{Troxel:2014dba,Joachimi:2015mma} for a review). 
%We mostly summarize known results from previous literature. A comprehensive review is e.g. \cite{Troxel:2014dba} \cit. 
There will be a few novel results in this section related to the fact that we do not take the flat-sky approximation, which is commonly employed in these studies (see for instance \cite{Blazek:2011xq}).  There are a few cases where this approximation should be dropped: one case is when looking at the imprint of local-type primordial non-Gaussianity on intrinsic alignments and the other is when looking at the signature of primordial gravitational waves, which is also the focus of this work. In both cases, the motivation for dropping the flat sky approximation is that most of the interesting signature is indeed at the largest scales, where the approximation breaks down. We will therefore present all our results, this section included, in the full-sky regime. 

Let us first of all define a three-dimensional field which describes galaxy shape perturbations
\begin{equation}
    g_{ij} (\vx,\tau) = \frac{I_{ij}(\vx,\tau) - \frac 13 \delta^{\rm K}_{ij}\, {\rm tr }[I_{\ell m}]}{{\rm tr }[I_{\ell m}]} \, ,
\end{equation}
where $I_{ij}$ is the symmetric second-moment tensor describing the intrinsic emissivity of a galaxy\footnote{While $I_{ij}$ is the proper intrinsic galaxy shape field, and $g_{ij}$ its perturbations, with a small abuse of terminology we will call $g_{ij}$ itself the galaxy shape field and galaxy shape power spectrum its two-point correlation function in Fourier space. In literature, $g_{ij}$ is also called the ``shear'' field, unifying terminologies with the weak lensing quantities.}.
The formation of a galaxy  must be determined by all sorts of physical processes taking place in the finite sized region of matter from which it originates, through a period of time which likely spans several decades of expansion. 
%For instance, it is reasonable to assume that some features of the early density field might determine the shape of a galaxy, in a similar way as peaks determine halos in the late time\mb{refine}. 
%It is certainly reasonable to assume that, at each moment in time, the shape of a galaxy is affected mostly by the local distribution of matter, where by local we mean some finite sized region, with scale $R_*$, typically larger than the size of the galaxy itself. %In this sense, this is a very similar approach as it is done in galaxy clustering where the observable is the scalar galaxy number density field. 
%In our treatment, we follow the approach of \cite{Schmidt:2015xka} and we restrict to studying scales larger than $R_*$. 
Assuming that gravity is the only force at play, we expect this process to be determined by perturbations of the gravitational potential, or rather its second derivative, $\partial_i\partial_j \Phi$, since the equivalence principle states that the  leading  locally  observable gravitational  effect is  given  by  second  derivatives  of  the  metric  tensor\footnote{As we will show in the next section, tensor perturbations of the metric also affect galaxy shapes, but sub-dominantly. We will therefore neglect them for now.}.  We therefore decompose $\partial_i\partial_j \Phi$ into two parts:  its trace, i.e. matter over-density field $\delta$, and the trace-free tidal tensor field
%The conventional way of decomposing the symmetric tensor $\partial_i\partial_j \Phi$ is to divide it into a scalar quantity, the matter over-density field $\delta$, which is connected to $\nabla^2 \Phi$ through the Poisson equation, and the trace-free tidal tensor field
\begin{equation}
    K_{ij} = \frac{1}{4\pi G \bar\rho a^2}\left[\partial_i\partial_j - \frac 13 \delta_{ij}\nabla^2\right]\Phi = \mathcal D_{ij}\delta \, ,
\end{equation}
where $\bar\rho$ is the mean energy density in the Universe, $a$ the scale factor and
\begin{equation}
    \mathcal D_{ij} \equiv \frac{\partial_i\partial_j}{\nabla^2} - \frac 13 \delta_{ij} \, .
\end{equation}
We should therefore expect that the galaxy shape field $g_{ij}$ can be expanded as a spatially local\footnote{While the expansion can be written as local in space, it is however non-local in time, as $g_{ij}$ depends on the full past history of $K_{ij}$ and $\delta$. A more appropriate definition would be
\begin{equation}\label{eq:func}
    g_{ij}(\bx,\eta) = \mathcal F\bigr[K_{ij}(\bx_{\rm fl}(\eta')), \delta(\bx_{\rm fl}(\eta'))\bigr]\, ,
\end{equation}
where $\bx_{\rm fl}(\eta')$ is the fluid trajectory from initial to final time, being therefore $\eta'>\eta$, and it shows explicitly that $g_{ij}$ at time $\eta$ depends on the past history of the trajectory. As shown in \cite{Vlah:2019byq}, the dependence on the fluid trajectory arises already at second order in the expansion Eq. \eqref{eq:func}. For the model we will consider later on, we neglect this dependence, and we reserve a more complete treatment for future work.}
functional of $\delta$ and $K_{ij}$,
\begin{equation}\label{eq:func1}
    g_{ij}(\bx,\eta) = \mathcal F\bigr[K_{ij}(\bx,\eta), \delta(\bx,\eta)\bigr] \, .
\end{equation}
Any deviation from the locality-in-space assumption enters as higher-order derivatives of $\partial_i\partial_j\Phi$, such as $\nabla^2 \delta$ and $\nabla^2 K_{ij}$, at the scale $R_*$, which is the size of the initial matter overdensity originating the galaxy. In the context of galaxy clustering, this scale is usually associated to the typical size of the halo hosting the galaxy, which is its Lagrangian radius. In this work, we are interested in large-scale correlations among galaxy shapes, we therefore neglect higher-order corrections.
%\footnote{In the case of galaxy shapes, we expect that neglecting terms sourced at higher-orders in derivatives of $\Phi$ might be more delicate than in the case of galaxy clustering, just because the shape of galaxies is very sensitive to the 3-dimensional environment surrounding it. Numerical studies confirm this expectation \cit.}. 
In order to make sure that perturbations smaller than $R_*$ do not affect the intrinsic alignments of galaxies, we smooth the tidal field $K_{ij}$  with a multiplicative window function in Fourier space,
\begin{equation}
    K_{ij,R}(k) = \left[\frac{k_i k_j}{k^2} - \frac 13\delta^K_{ij}\right] W_R(k) \delta(k) \, ,
\end{equation}
being $W_R(k)= e^{- k^2 R^2 /2}$
a Gaussian filter and similarly for $\delta$. In this work, we will assume a Gaussian smoothing in Fourier space with $R_* = 1$ Mpc/h, which would correspond to a halo of about $M \sim \mathcal O(1) \times 10^{11} M_\odot$. We will suppress the subscript $R$ with the understanding that the tidal field is always smoothed\footnote{While a more refined smoothing should be considered, our results would not change qualitatively and are easily extended to more realistic scenarios. In particular, the choice of scale $R$ does not affect significantly the results of this analysis, but for one of the (subdominant) contributions to the galaxy shape power spectrum, which we will discuss more in detail.}.

\subsection{Linear alignment model}
In order to make progress, we need now to specify how $g_{ij}$ responds to changes in long-wavelength perturbations of $K_{ij}$ and $\delta$, or in other words to specify the form of the functional in Eq. \eqref{eq:func1}.
The most unassuming and complete way of implementing this would be to use an effective field theory approach, as done in \cite{Vlah:2019byq}, including all allowed operators in the expansion. This method would allow also to implement corrections from higher-order derivative terms and the non-locality in time in a straightforward way. For the present analysis, we will instead focus on a specific model, which assumes that $g_{ij}$ responds (only) linearly to $K_{ij}$
\begin{align}\label{eq:gammaexp}
    g_{ij}(\bx,\tau) \simeq&\, b_K K_{ij}(\bx,\tau) \, , %+ b_{\delta K}^\gamma K_{ij}(\bx,\tau)\delta(\bx,\tau)+b_{K^2}^\gamma \left[K_{i\ell}(\bx,\tau)K^{\ell}_j(\bx,\tau)-\frac 13 \delta_{ij} K^2 (\bx,\tau)\right]+...%  \nonumber\\
    %&+ b^\gamma_{K^2K}K^2 K_{ij} + b^\gamma_{\delta^2 K} K_{ij} \delta^2 + b^\gamma_{\delta K^2} \left[K_{i\ell}(\bx,\tau)K^{\ell}_j(\bx,\tau)-\frac 13 \delta_{ij} K^2 (\bx,\tau)\right]\delta
    %+... 
\end{align}
where 
%the ellipses indicate higher order terms in $\delta$ and $K_{ij}$ as well as non-local terms, whose responses in this model are assumed to be small. T
the parameter $b_K$ is the galaxy shape linear bias and it has the same interpretation as bias parameters of the local number density of galaxies as the response of the galaxy shape to a change in the local value of the tidal tensor $K_{ij}$. 
The Linear Alignment (LA) model was introduced early on by \cite{Catelan:2000vm} and it is frequently used when dealing with populations of red galaxies. The idea behind it is that the galaxy ellipticity is driven by that of the halo hosting it, and that for small enough perturbations on large enough scales, the response would indeed be linear as in the case of linear galaxy biasing \cite{Hirata:2004gc} \footnote{It was also argued by \cite{Hirata:2004gc} that spiral galaxies would respond to $K_{i\ell} K^{\ell}_j$, in what is called the tidal torque model, hence breaking this assumption. Simulations also show that galaxy and star formation physics can erase almost completely the initial alignment, therefore breaking the assumption that gravitational collapse is the only physics at play (see \cite{Troxel:2014dba} and references therein).}.  Observations have so far shown good agreement with this model for Luminous Red Galaxies (LRG) at redshift $z\sim 0.3$ \cite{Mandelbaum:2005hr, Joachimi:2010xb,Hirata:2007np,Blazek:2011xq,Singh:2014kla}. Different types of galaxies do not show a similar agreement \cite{Mandelbaum:2009ck,Heymans:2013fya}. 
Even within this model, there are a few subtleties that need to be clarified. For instance, a simplified scenario might be that the intrinsic alignment is imprinted at some early redshift $z=z_P$ during matter domination and it stays frozen until the observation time $z=z_O$. This implies that the amplitude of the response should depend explicitly on $z_P$, so that $b_K \propto D(z_P) / D(z_O)$, where $D(z)$ is the linear growth factor. Unless the galaxy is very old, this is usually a factor of order unity. For instance, in the case of the intrinsic alignment observed in LRG galaxies at redshift $z_O = 0.3$, assuming that the alignment was imprinted while the galaxy was forming, i.e. around $z=2$, we would have $D(z_P=2) / D(z_O=0.3)\sim 0.5$. For the present analysis, these factors would not change significantly our final results, therefore we will just assume $z_P \equiv z_O$.

\paragraph{Density-weighting} An important point to make is that the galaxy shape field should be generically weighted by the galaxy number density field, since the information on the shapes comes necessarily from light emitted by an observable galaxy \cite{Blazek:2011xq}. We therefore work with the weighted field $\tilde g_{ij} = g_{ij}(1+\delta_g)$, where $\delta_g$ is the galaxy number density field contrast, which has its own expansion in terms of $\delta$ and $K_{ij}$
\begin{equation}\label{eq:deltagexp}
    \delta_g(\bx,\tau) = b_\delta \delta (\bx,\tau) + 
    b_{\delta^2} \delta^2(\bx,\tau) + b_{K^2} K^2(\bx,\tau)+ 
    ...\, ,
\end{equation}
where $K^2$ is the square of $K_{ij}$ and the ellipses again indicate higher-order terms in $\delta$. The criterion for truncating the expansion is that we want to include all terms up to $\mathcal O(P_\delta^2)$, where $P_\delta$ is the linear matter power spectrum. The density-weighted galaxy shape expansion therefore reads
\begin{align}\label{eq:exptilde}
    \tilde g_{ij} =&\, b_K K_{ij} + b_\delta b_K \delta K_{ij} + b_\delta^2 b_K \delta^2 K_{ij} + b_{K^2} b_K K^2 K_{ij} + ... \, ,
    %+b_{K^2}^\gamma \left[K_{i\ell}K^{\ell}_j-\frac 13 \delta_{ij} K^2\right] +(b^\gamma_{\delta K^2}+ b_\delta^g b_{K^2}^\gamma) \left[K_{i\ell}K^{\ell}_j-\frac 13 \delta_{ij} K^2\right] \delta \nonumber\\
    %& + (b^\gamma_{\delta^2 K}+b_\delta^g b_{\delta K}^\gamma+b_K^\gamma b_{\delta^2}^g)  K_{ij}\delta^2  + (b^\gamma_{K^2K}+ b^\gamma_{K} b^g_{K^2}) K_{ij} K^2...\,,  
\end{align}
where 
%we keep terms of order such that $\langle \tilde\gamma \tilde \gamma\rangle$ is proportional to up to $P_\delta^2$ and 
we suppressed dependence on $\bx$ and $\tau$ to avoid clutter. 
%The expansion above shows a high degree of degeneracy among bias parameter, although they have different physical interpretation. 
For similar reasons as argued for the galaxy shape field, the galaxy density field should also be smoothed on some scale $R'$. A usual choice is to use a top-hat smoothing in Fourier space, with $R'$ being again the Lagrangian radius of the halo/galaxy. Here we choose, for the sake of simplicity, to just use the same smoothing as for the galaxy shape field, $W_R(k) = e^{-k^2 R^2/2}$ at the same scale $R_*=R'$. In the context of halo clustering, bias parameters from the galaxy density field expansion might be predicted, for instance, using excursion set approaches combined with peak statistics (see \cite{Desjacques:2016bnm} for a review), but  bias parameters related to the tidal field $K_{ij}$ are known to be difficult to predict in these models \cite{Desjacques:2017msa}. 
%On the other hand, if we follow an effective approach, we should write an expansion in terms of bare bias parameters for each operator and then renormalise them, since most of the terms of the expansion are operators computed in the same position in space. The effective expansion in terms of bare parameters then reads
%\begin{align}\label{eq:exptilde}
%    \tilde\gamma_{ij} =&\, c_K K_{ij} + c_{\delta K}  K_{ij}\delta+c_{K^2} \left[K_{i\ell}K^{\ell}_j-\frac 13 \delta_{ij} K^2\right]  \nonumber\\
%    & +c_{\delta K^2} \left[K_{i\ell}K^{\ell}_j-\frac 13 \delta_{ij} K^2\right] \delta+ c_{\delta^2 K} K_{ij}\delta^2  + c_{K^2 K} K_{ij} K^2...\,,  
%\end{align}
%Now need to renormalize.

\paragraph{Projection in the sky.}

Until now, we have expressed the galaxy shape $g_{ij}$ in terms of a 3D field as the physical processes that can contribute to it are explicitly dependent on all three directions. However, observations of galaxy shapes are made through 2D images from galaxy surveys, which are the projection of $g_{ij}$ on the sky. We therefore define the density-weighted intrinsic shape field as\footnote{In literature, the projection $\tilde \gamma$ has been defined with a superscript ``IA'' to distinguish the intrinsic alignments from the  gravitational lensing shear field, being the total projected shape field the sum of the two. Here we do not consider contributions from lensing, hence there is no ambiguity of definitions.}
%\begin{equation}
%    \tilde \gamma_{ij} \equiv \mathcal P_i^\ell\, \mathcal P^m_j\, \tilde g_{\ell m},
%\end{equation}
\begin{equation}
    \tilde \gamma_{ij} (\hat \bn)= \int dz \, \frac{d N}{d z} \, \mathcal P_i^\ell\, \mathcal P^m_j\, \tilde g_{\ell m}(\chi(z)\hat \bn,\eta(z)) \, ,
\end{equation}
being $\mathcal P_{ij} = \delta_{ij}-\hat n_i \hat n_j$ the projection operator onto the sky\footnote{Written as it is, $\tilde \gamma$ is not a trace-free quantity. A proper definition would be
\begin{equation}
    \tilde \gamma_{ij} (\hat \bn)= \int dz \, \frac{d N}{d z} \,  \left(\mathcal P_i^\ell\, \mathcal P^m_j-\frac 12 \mathcal P_{ij}\mathcal P^{\ell m}\right)\, \tilde g_{\ell m}(\chi(z)\hat \bn,\eta(z)) \, ,
\end{equation}
where the trace is explicitly subtracted. It turns out that when decomposing into $\pm 2$ spin functions as done below in Eq. \eqref{eq:pm2}, the second term in brackets gives zero because of the properties of the unit vectors $m_\pm$.}, 
%Depending on the specific imaging survey observing galaxy shapes, these are observed at different red-shifts with different precisions. Therefore, we should really consider the projected field integrated over a red-shift range
$d N/d z $ the redshift distribution of a specific imaging survey, $\chi(z)$ the comoving distance out to redshift $z$ and $\eta(z)$ the conformal time. In what follows, since we are mostly interested in order of magnitude estimations, we simplify our calculations considering a single observed redshift $d N/d z = \delta_{\rm D}(z-z_O)$.
\paragraph{Harmonic decomposition.} The projected field $\tilde \gamma_{ij}$ is a traceless 2-tensor on the sphere. It is therefore natural to compute its angular correlations expressed in terms of multipole moments. In order to do that, we need to apply spin-lowering and -raising operators on $\tilde \gamma_{ij}$ to convert it into a scalar in the sky, $\tilde\gamma$. We give details on this procedure in Appendix \ref{app:scalars}. The harmonic sphere coefficients of $\tilde \gamma$ are given as 
\ba\label{eq:spharm}
a_{\ell m}^{\tilde \g} = \sqrt{\frac{(\ell-2)!}{(\ell+2)!}} \int d\Omega  \, Y^*_{\ell m}(\vnhat) \, \tilde \gamma(\vnhat,\vk) \, ,
\ea
where $Y_{lm}$ is the spherical harmonic function. The integral over the angle can be performed analytically following the identity \cite{Schmidt:2012ne}
\ba \label{harmonic_eq}
 \int d\Omega  \, Y^*_{\ell m}(\vnhat) (1-\mu^2)^{|r|/2}\, e^{ i r \phi} e^{i \mu x} = i^{r+\ell} \sqrt{4 \pi (2 \ell + 1)} \sqrt{\frac{(\ell+|r|)!}{(\ell-|r|)!}} \frac{j_\ell(x)}{x^{|r|}} \, \delta_{m r}\, ,
\ea
where $j_\ell(x)$ are the spherical Bessel functions of the first kind. As a result, the operators $\hat Q_n(x)$ act on the Bessel functions generating transfer functions 
\begin{align}
    F^{E |r|}_\ell(x) &\equiv {\rm Re}\left[\hat Q_r (x)\right]\,\frac{j_\ell(x)}{x^{|r|}}\\
    F^{B |r|}_\ell(x) &\equiv {\rm Im}\left[\hat Q_r (x)\right]\,\frac{j_\ell(x)}{x^{|r|}}\, ,
\end{align}
being $r=0,\pm 1,\pm2$, which are functions of $\ell$ and $x$ and can be found in Appendix \ref{app:scalars}.  We can now express the coefficients $a_{lm}$ in terms of E and B modes as
\ba\label{eq:almeb}
a^{ E}_{lm} =\:& \frac12\left(a_{lm} + a^{*}_{lm} \right) \vs
a^{B}_{lm} =\:& \frac1{2i}\left(a_{lm} - a^{*}_{lm} \right)\, ,
\ea 
and consequently define the power spectra as 
\be
C^{XX'}_{\ell} = \frac{1}{2 \ell +1} \sum_m \int \frac{d^3 k}{(2 \pi)^3} \int \frac{d^3 k'}{(2 \pi)^3} \langle a_{\ell m}^{ X} a_{\ell m}^{X'*} \rangle \, ,
\ee
where $X=E,B$. We provide the full expressions for the harmonic coefficients $a_{\ell m}$ for each term of Eq. \eqref{eq:pmscalar} in Appendix \ref{app:scalars}.    %The last step is to integrate over a certain galaxy redshift distribution. 
%Here we choose...
\subsection{Galaxy shape power spectrum}

We now have all the ingredients to compute the two-point correlation of intrinsic galaxy shapes at large scales. Within the LA model that we consider, there are five terms in total that contribute up to order $\mathcal O(P^2)$\footnote{These calculations have been performed in earlier analyses (see for instance \cite{Blazek:2015lfa}), but always in the flat sky approximation. We have checked that our results coincide in that limit, although ours are valid also at the largest scales, where the flat sky approximation breaks down.}. For better intuition, we can momentarily drop indices and schematically write down the galaxy shape power spectrum as
\begin{align}\label{eq:intuit}
    \langle \tilde \gamma\, \tilde \gamma \rangle = &\, b_K^2\,  \left\langle K\, K \right\rangle + 2 \,b_K^2\, b_\delta\,  \left \langle K\, (K*\delta) \right\rangle  + b_K^2 b_\delta^2 \langle (K*\delta)\,(K*\delta)\rangle + \nonumber\\
    &+2\, b^2_K\, b_{\delta^2}\, \langle K\, (K*(\delta*\delta)) \rangle+ 2\, b_K^2\, b_{K^2}\,\langle K\, (K*(K*K)) \rangle + \mathcal O(P_\delta^3)\, ,
\end{align} 
where $*$ indicates convolution in Fourier space. Let us consider each of these terms separately. 
\paragraph{ The $\langle K K \rangle$ term.} This is the leading order term coming from combining the tidal shear tensor $K_{ij}$ with itself. At this order, no B-mode is sourced, therefore we have
\begin{align}
 C_\ell^{EE, (KK)} =&\,  b_K^2 \,\, \int \frac{dk}{2\pi}\, k^2\,  \frac{(\ell-2)!}{(\ell+2)!} |F^{E0}_\ell(x)|^2\, P^{\rm 1L}_\delta(k) \, ,\\
 C_\ell^{BB, (KK)} =& 0 \, ,
\end{align}
where here $P^{\rm 1L}_\delta(k)$ is matter power spectrum up to one-loop defined as
\begin{equation}
    P^{\rm 1L}_\delta(k) = P_\delta(k) + P^{(22)}(k) + P^{(13)}(k) \, ,
\end{equation}
being $P_\delta(k)$ the linear matter power spectrum and
\begin{align}
    P^{(22)}(k) & = 2\int_{\vq} P_\delta(q) P_\delta(|\vk-\vq|)F_2^2(\vq, \vk-\vq)\, ,\\
    P^{(13)}(k) & = 6 P_\delta (k) \int_{\vq}  P_\delta(q) F_3(\vk,\vq,-\vq)\, ,
\end{align}
are the one-loop standard perturbation theory (PT) corrections to the linear matter power spectrum, where the superscript $^{(n)}$ indicates the order in PT, and
\begin{equation}
    F_2(\vk_1,\vk_2) = \frac 57+\frac 12 \frac{\vk_1\cdot\vk_2}{k_1 k_2}\left(\frac{k_1}{k_2}+\frac{k_2}{k_1}\right) + \frac 27 \left(\frac{\vk_1\cdot\vk_2}{k_1 k_2}\right)^2
\end{equation}
and $F_3$, whose expression is found in the comprehensive review \cite{Bernardeau:2001qr}, are the PT kernels. Each linear matter power spectrum is smoothed as indicated in the previous paragraph. In previous work \cite{Blazek:2011xq, Blazek:2015lfa}, the nonlinear matter power spectrum (e.g. using halofit) was used in order to extend the validity of this contribution to smaller scales. Here we are interested only in the large scales, therefore we stick to the one-loop result. 

\paragraph{ The $\langle K (K*\delta) \rangle$ term.} Gravitational mode-coupling sources a non-zero bispectrum at one-loop order in perturbation theory (PT). The matter density at second-order in PT reads
\begin{equation}
    \delta^{(2)}(\vk) = \int_{\vq} F_2(\vq,\vk-\vq)\delta^{(1)}(\vq)\delta^{(1)}(\vk-\vq)\, .
\end{equation}
Consequently, we get a contribution from three bispectra of the type
\begin{equation}
    \langle K (K*\delta) \rangle = \langle K^{(2)} (K^{(1)}*\delta^{(1)}) \rangle + \langle K^{(1)} (K^{(2)}*\delta^{(1)}) \rangle + \langle K^{(1)} (K^{(1)}*\delta^{(2)}) \rangle\, ,
\end{equation}
where $K^{(2)} = \mathcal D_{ij} \delta^{(2)}$. Computing these terms in harmonic space we get
\begin{align}\label{eq:f2}
    C_\ell^{EE,(K K\delta )} &= b_K^2\, b_{\delta}\, \int \frac{dk}{2\pi}\, k^2\frac{(\ell-2)!}{(\ell+2)!} |F^{E0}_\ell(x)|^2  \left[ S_{F_2}(k) + R(k)P(k)+ \frac{29}{105}\sigma^2 P(k) \right] \\
 C_\ell^{BB, (K K\delta )} &= 0 \, ,
\end{align}
where notice that in this case also there is no B-mode sourced. The functions $S_{n}(k)$ and $R(k)$ are defined as
\begin{align}
    S_n(k) & = \frac{k^3}{(2\pi)^2}\int_0^\infty dr \int_{-1}^{1}d\mu\, P(r k) P( k\sqrt{1+r^2-2 r \mu}) \tilde S_n(r,\mu)\\
    R(k) & = \frac{k^3}{(2\pi)^2}\int_0^\infty dr\,  P(r k) \tilde R(r)\, ,
\end{align}
where we provide more details on the calculation, along with the kernels $\tilde S_n$ amd $\tilde R$ in Appendix \ref{app:scalars}. Similarly to what done in \cite{Blazek:2015lfa}, in $R(k)$ we have subtracted the $k\rightarrow 0$ piece and added it back to the third term, which renormalizes the linear shape bias as
\begin{equation}\label{eq:bkcorr}
    b_K^2 \rightarrow b_K^2 \left(1 + \frac{58}{105}\sigma^2 b_\delta\right)\, .
\end{equation}
%In terms of E and B decomposition, the E and B auto-correlations are\mb{show first piece by piece}
%\begin{align}
% C_\ell^{EE} =&\,  b_K^2 \,\frac{2}{\pi}\, \int dk\, k^2\, \left[ \frac{(\ell-2)!}{(\ell+2)!} |F^{E0}_\ell(x)|^2 %\left(\frac 14 P_\delta(k) + b_{\delta}^2 S_0(k)\right) + \right.\nonumber\\
% &\left.+\frac{1}{\ell^2+\ell-2}\, |F^{E1}_\ell(x)|^2 S_1(k)+  |F^{E2}_\ell(x)|^2 S_2(k)\right]\\
% C_\ell^{BB} &= b_K^2 b_\delta^2 \frac{2}{\pi} \int dk\, k^2\, \left[  \frac{1}{\ell^2+\ell-2}\, |F^{B1}_\ell(x)|^2 S_1(k)+  |F^{B2}_\ell(x)|^2 S_2(k)\right],
%\end{align}
%where we define the $S$ functions in appendix \ref{}. \mb{consider splitting in order P and order P2, add F2 piece}.
\paragraph{The $\langle (K*\delta) (K*\delta) \rangle$ term.} This term involves the correlation of the tidal shear field $K$ with the galaxy density field $\delta_g$,
\begin{align}\label{eq:bbscalars}
    C_\ell^{EE,(K \delta)^2} &= b_{K}^2\, b_\delta^2  \int \frac{dk}{2\pi}\, k^2 \left[ \frac{(\ell-2)!}{(\ell+2)!} |F^{E0}_\ell(x)|^2 S_0(k) + \frac{1}{\ell^2+\ell-2}\, |F^{E1}_\ell(x)|^2 S_1(k)+  |F^{E2}_\ell(x)|^2 S_2(k)\right]\\
    \label{eq:bbconvol}
    C_\ell^{BB,(K \delta)^2} &= b_{K}^2 \, b_\delta^2  \int \frac{dk}{2\pi}\, k^2 \left[  \frac{1}{\ell^2+\ell-2}\, |F^{B1}_\ell(x)|^2 S_1(k)+  |F^{B2}_\ell(x)|^2 S_2(k)\right] \, .
\end{align}
%where
%\begin{align}
%    \hat S_0(k) &= S_0(k) - \frac 15 \sigma^2\\
%    \hat S_1(k) &= S_1(k)\\
%    \hat S_2(k) &= S_2(k) - \frac{1}{15}\sigma^2
%\end{align}
 The functions $S_0$, $S_1$ and $S_2$ involve the convolution of two power spectra, 
\begin{equation}
    \int_{\vq} P(q)P(|\vk-\vq|)\, ,
\end{equation}
which allows for the transfer of power from E modes to B modes, as noted first in \cite{Hirata:2004gc}. In the context of galaxy clustering, analogous terms arise in the computation of the galaxy power spectrum and are known to be very sensitive to the smoothing scale $R$ (see \cite{McDonald:2006mx,Assassi:2014fva,Desjacques:2016bnm} for more discussion). Moreover, these integrals go to a constant as $k\rightarrow 0$, therefore acting as a shot-noise term at large scales. In our case, this implies an $\ell$-independent contribution to the $C_\ell^{X (\delta K)^2}$ power spectrum at low $\ell$. In analogy with the Fourier space case, we subtract the $\ell=2$ contribution and therefore show results from $\ell \geq 3$. The $\ell=2$ contribution would need to be then added to the total shot-noise power spectrum, which we do not show here (but see Section  \ref{sec:chiral} for a short discussion). %We explain in more details  how we subtract the shot-noise contribution in the Appendix \ref{app:scalars}. 
%Upon checking the limit of the whole integrand for $k\rightarrow 0$, we realize that the transfer functions $F^{E0}_2$, $F^{E2}_2$ and $F^{B1}_2$ do not vanish. We find the following limits
%\begin{align}
%   \frac{(\ell-2)!}{(\ell+2)!} |F^{E0}_\ell(x)|^2 S_0(k) &\stackrel{k\rightarrow 0}{\longrightarrow} \frac{1}{1125} \Sigma^2 \delta^K_{\ell2} \\
%    |F^{E2}_\ell(x)|^2 S_2(k) &\stackrel{k\rightarrow 0}{\longrightarrow} \frac{16}{375} \Sigma^2 \delta^K_{\ell2} \\
%    \frac{1}{\ell^2+\ell-2}|F^{B1}_\ell(x)|^2 S_1(k) &\stackrel{k\rightarrow 0}{\longrightarrow} -\frac{4}{375} \Sigma^2 \delta^K_{\ell2},
%\end{align}
%where $\Sigma^2 = 1/2\pi^2 \int {\rm d} q\, q^2 P^2(q)$. Therefore, the quadrupole does not vanish for $k\rightarrow 0$. 
\begin{figure}[ tbp]
	\centering
	\includegraphics[width=0.495\textwidth]{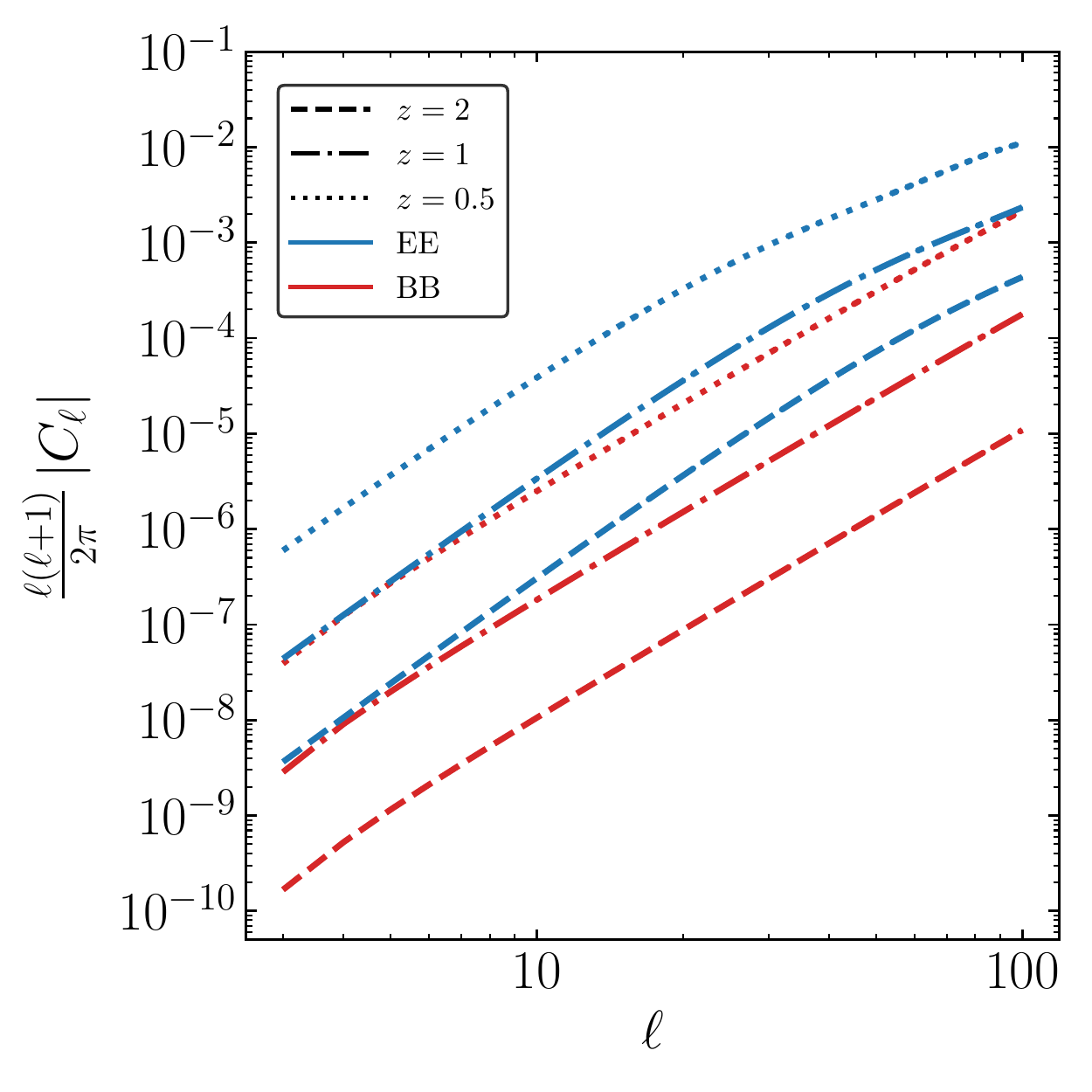}
	\includegraphics[width=0.495\textwidth]{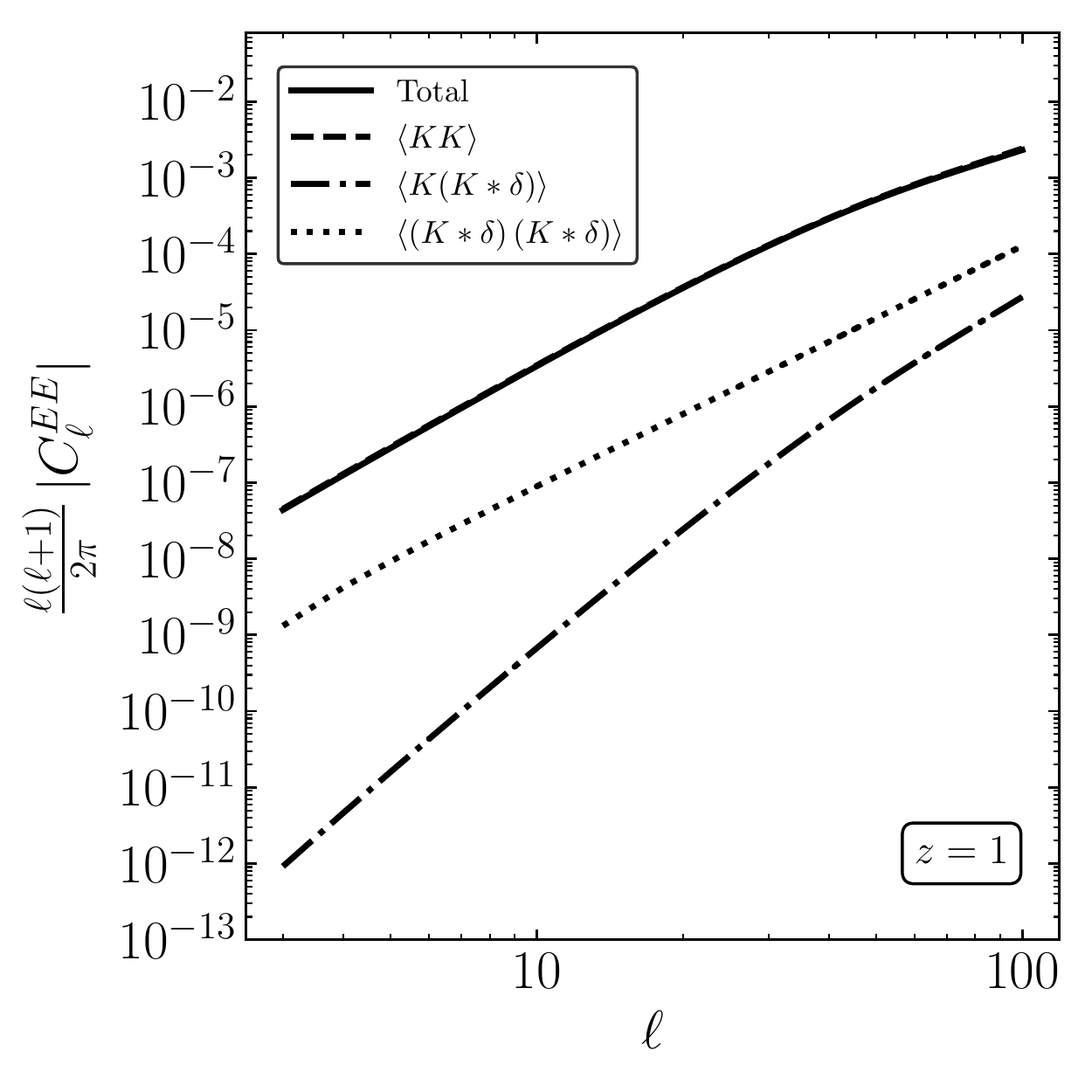}
	\caption{Left panel: EE and BB density-weighted galaxy shape power spectrum at three different redshifts as generated in the LA model by the tidal shear field $K_{ij}$. Right Panel: all contributions to the EE density-weighted galaxy shape power spectrum at redshift $z=1$.}
	\label{fig:scalarterms}
\end{figure}
\paragraph{The $\langle K (K*(\delta*\delta) \rangle$ and the $\langle K (K*(K*K) \rangle$ terms.} These terms also renormalize the linear bias. We therefore add them to Eq. \eqref{eq:bkcorr} and get the following renormalization
\begin{equation}
    b_K^2 \rightarrow b_K^2 \left[1 + \sigma^2\left(\frac{58}{105} b_\delta -  2 b_{\delta^2} + \frac{28}{15}  b_{K^2}\right)\right]\, ,
\end{equation}
%Clearly, the amplitude of tidal bias parameters determine which terms are relevant. While the EE power spectrum has been measured and found in agreement with the leading term, the BB power spectrum is still unobserved so it is not possible to say yet much about bias parameters beyond $b_K$. 
with more details on the calculation in Appendix \ref{app:scalars}.

In Figure \ref{fig:scalarterms} we show the total galaxy shape power spectrum in the LA model at three different redshifts $z=0.5, 1, 2$ and a comparison of all the terms at redshift $z=1$. For these figures, we have simply chosen $b_K = b_\delta = 1$ and all the power spectra are multiplied by the linear growth factor $D^2(z)$. A more detailed discussion about bias parameters is in order: according to what found in LRG observations, $b_K = - C_1 \Omega_m D(z=0)/ D(z_O)$, where $C_1=0.12$ and the growth factor needs to be normalized to be $(1+z) D(z) =1$ during matter domination \cite{Blazek:2011xq}. As for the linear bias, this was measured for the same dataset to be of order $b_\delta \simeq 2$. While for the linear bias we just choose $b_\delta=1$ at all redshifts, it is generically higher at higher redshifts. We have employed a flat $\Lambda$CDM cosmology with $h=0.7$, $\Omega_m=0.3$ and  $\sigma_8=0.85$ and the linear matter power spectrum is computed using the CLASS code \cite{Blas:2011rf}. We are only showing large scales up to $\ell =100$, which roughly corresponds to $k\sim 0.03$ h/Mpc at redshift $z=2$, but we have developed a freely-available code\footnote{\textrm{https://gitlab.com/mbiagetti/tensor\textunderscore fossil}} for the fast computation of these power spectra at high-$\ell$, showing its accuracy at low $\ell$ in Figure \ref{fig:approx}. We give more details about these methods in Appendix \ref{app:scalars} and \ref{app:approx}. %\mb{point out that our smallest scale at $\ell=100$, should correspond to $\sim \ell/\chi(z)$, at redshift $z=2$ is $\sim0.03$ h/Mpc.}

\section{Primordial gravitational waves and intrinsic alignments}\label{sec:tensors}

In the previous section,  we assumed that galaxy shapes respond linearly to changes in the local tidal shear field, $K_{ij}$. It has been pointed out early on \cite{Dodelson:2003bv,Dodelson:2010qu} that $K_{ij}$ is not the only source of the external tidal field correlating galaxy shapes at large scales: tensor perturbations in the metric contribute as well. The inflationary scenario indeed predicts the generation of such propagating tensor modes, known as primordial gravitational waves. Recent efforts \cite{Schmidt:2012ne,Schmidt:2012nw,Schmidt:2013gwa} have elaborated on how to consistently compute what is the imprint of primordially generated tensor perturbations in the late universe and specifically their impact in the local distribution of matter.  In this section, we summarize these findings and show how galaxy shapes respond to primordial gravitational waves.

\subsection{Tensor perturbations from inflation}

Let us start by defining transverse and traceless tensor perturbations $h_{ij}$ around a flat FLRW metric
\begin{equation}
    d s^2 = a^2(\eta)\left[-d\eta^2+ (\delta_{ij}+h_{ij})dx^i dx^j\right] \, ,
\end{equation}
where $a(\eta)$ is the scale factor in conformal time $\eta$. These are generically sourced by an inflationary scenario \cite{Starobinsky:1979ty}. Scalar perturbations are also present and sourced during inflation, but we neglect them for the time being. The tensor field $h_{ij}$ can be decomposed into Fourier modes of two polarization states
\begin{equation}
    h_{ij}(\bk) = \sum_{s=R/L} \epsilon_{ij}^s(\hat\bk)h_s(\bk) \, ,
\end{equation}
where, for the purpose of what follows, we choose to use chiral polarizations states defined through 
\begin{align}\label{definition_LR}
\epsilon^R_{ij} &= \epsilon^+_{ij}+ i \epsilon^{\times}_{ij} \mbox{ },\\
\epsilon^L_{ij} &= \epsilon^+_{ij} - i \epsilon^{\times}_{ij} \mbox{ },\\
h_{R/L} &= \frac{h_+ \mp i h_\times}{2}\mbox{ },
\end{align}
where $\epsilon^{+/\times}_{ij}$ and $h_{+/\times}$ define the usual two linear independent polarizations of primordial gravitational waves. 
%The two chiral polarization states satisfy the following relations
%\begin{align} 
%\epsilon^L_{ij}(\bk)\epsilon_{L}^{ij}(\bk)&= \epsilon^R_{ij}(\bk)\epsilon_{R}^{ij}(\bk)= 0 \mbox{ }, \nonumber\\
%\epsilon^L_{ij}(\bk)\epsilon_{R}^{ij}(\bk)&= 4 \mbox{ }, \nonumber\\
%\epsilon^{R*}_{ij}(\bk)&= \epsilon^L_{ij}(\bk) \nonumber \mbox{ },\\ 
%k_l \epsilon^{mlj} \mbox{ } {{\epsilon_{(s)}}_{j}}^i(\bk) &= - i \mbox{ } \lambda_s k \mbox{ }\epsilon_{(s)}^{im}(\bk) \nonumber \mbox{ ,}\\
%h^*_R(\bk) & = h_L(\bk) \label{relations_LR}\, ,
%\end{align}
%where $k_i$ is the $i$-th component of the momentum $\mathbf k$, $\lambda_R = +1$ and $\lambda_L = - 1$. Here $\epsilon^{ijk}$ is the Levi-Civita pseudo-tensor with three Latin indices. Also we recall that $s$ is the polarization index and not a tensor index. In Eqs. \eqref{relations_LR} Latin contractions are made with the $\delta_{ij}$.

The total primordial power spectrum of gravitational waves is then defined through
\begin{equation} \label{Powerh_def}
    \langle h_{ij}(\bk,\eta)h^{ij}(\bk',\eta')\rangle = (2\pi)^3 \delta_D(\bk+\bk') P_h(k, \eta, \eta') \, .
\end{equation}
Notice that we can also define the amplitude of the single chiral mode as
\begin{equation}
    \langle h_{R/L}(\bk,\eta)h_{R/L}(\bk',\eta')\rangle = (2\pi)^3 \delta_D(\bk+\bk') P_{R/L}(k, \eta, \eta') \, .
\end{equation}
In models of inflation where primordial gravitational waves are unpolarized ($P_R = P_L$), we get
\begin{equation}
P_{R/L}(k, \eta, \eta') = \frac{P_h(k, \eta, \eta')}{8} \, ,
\end{equation}
where we have used Eq. \eqref{Powerh_def} and the fact that by definition $h_{R/L} = \epsilon^{ij}_{L/R} \, h_{ij}/4$.

In order to study the impact of these tensor modes at later time, we need to define a transfer function
\begin{equation}\label{eq:ttransfer_intro}
    h_{ij}(\mathbf{0},\eta) = T_h(\eta) h^{(0)}_{ij}(\mathbf{0}) \, ,
\end{equation}
where we have used the notation $h_{ij}(\bx, \eta=0) \equiv h^{(0)}_{ij}(\mathbf{x})$, $\eta= 0$ denoting the time at the end of inflation.
In matter domination, it takes the simple form
\begin{equation}\label{eq:ttransfer}
    T_h(\eta) = 3\frac{j_1(k_L\eta)}{k_L\eta} \, ,
\end{equation}
valid for a single Fourier mode $k_L$. In what follows, we use the full numerical transfer function for a LCDM universe, i.e. including radiation and $\Lambda$ domination phases, following \cite{Schmidt:2013gwa}.
The total tensor power spectrum at the end of inflation is parametrized as
\begin{equation} \label{P_tensors}
    P_h(\mathbf 0, k)= P_h(k) = \frac{2\pi^2}{k^3} r \mathcal A_s \, ,
\end{equation}
where $r$ is the tensor-to-scalar ratio, $A_s$ is the amplitude of scalar perturbations and we approximate the spectral index of tensor modes, $n_T = d\ln P_h / d\ln k \approx 0$, as it affects only negligibly our results. We will consider $r=0.1$ as a reference value in all our calculations.

\subsection{Primordial gravitational waves in the galaxy shape power spectrum}
%In constructing the effective expansion of Eq. \eqref{eq:exptilde}, we have neglected the fact that tensor modes generated during inflation provide a source for a pure tensor tidal field, which acts in a similar way as $K_{ij}$, but has primordial origin, on the galaxy shape field. Propagating gravitational waves are a generic prediction of the inflationary scenario, although their amplitude can vary across several orders of magnitudes depending on the inflationary model. Constraints on this amplitude are currently set by the Planck experiment to be at least $6\%$ smaller than their scalar counterpart, which was measured. 
The calculation of the imprint of tensor modes on intrinsic galaxy shapes is based on the essential fact that, generically, the leading locally observable effect of a long-wavelength perturbation $k_L$, of any kind, on a region of size much smaller than $1/k_L$ is an effective tidal field. For scalar perturbations, this is what generates the well-known $F_2$ kernel on the matter density field in standard perturbation theory \cite{Schmidt:2013gwa}. For tensor modes specifically, and in the context of galaxy shapes, it helps to use the Fermi normal coordinate (FNC) frame \cite{Fermi:1922,Manasse:1963zz}. In this frame, the metric $g_{\mu\nu}^F$ is Minkowski along the central geodesic passing through the center of mass of a given region of the Universe, with the relevant corrections of order $x^2_F$. It is possible to show that results obtained in these coordinates have a clear physical interpretation as corresponding to what a local freely falling observer moving along the central geodesic would measure (see e.g. \cite{Schmidt:2012ne}). 
%In particular, since in galaxies we deal with non-relativistic motions, the equations of motion of a mass test are completely described by the following $g_{00}^F$ component of the metric \cite{Schmidt:2012nw}
%\begin{equation} \label{metric_FNC}
%g_{00}^F = -1 + (\dot H + H^2) r_F^2 - 2 \Psi^F  \, ,   
%\end{equation}
%where $r_F^2 = \delta_{i j} x^i_F x^j_F$ and
%\begin{equation}\label{psi_F_cosmo}
%\Psi^F = -\frac{1}{4}
%\left({\ddot h}_{ij} + 2 H{\dot h}_{ij}\right)
%x_F^i x_F^j \, 
%\end{equation}
%is an effective peculiar potential sensitive to long scales tensor perturbations. From Eq. \eqref{psi_F_cosmo} it follows that the potential $\Psi^F$ gives rise to an external tidal force, causing the deformation of the galaxy shapes with respect to the rotational symmetry. For the purpose of our work it is convenient to work with the conformal time instead of cosmological time. So, the peculiar potential in conformal time reads 
%\begin{equation}
%\Psi^F = -\frac{a^{-2}}{4}
%\left({h''}_{ij} + a H{h'}_{ij}\right)
%x_F^ix_F^j \, ,
%\end{equation}
Using this framework, tensor perturbations source the following external tidal field at time $\eta$
\begin{align} \label{tidal_tens}
t_{ij} &= %\left( \partial_i \partial_j - \frac{1}{3} \delta_{ij} \partial^2\right)\Psi^F  = -\frac{a^{-2}}{2}
%\left({h''}_{ij} + a H {h'}_{ij}\right)\nonumber\\
%&=
-\frac{a^{-2}}{2} \left[T''_h(\eta) + aH T'_h(\eta)\right] h^{(0)}_{ij} \, ,
\end{align}
where the prime means derivative with respect to the conformal time and $T_h(\eta)$ is the transfer function of tensor perturbations from inflation as introduced in \eqref{eq:ttransfer_intro}. The dependence of the tidal field on time derivatives of the transfer function $T_h$ implies that it vanishes on superhorizon scales, as one would expect, and only tensor perturbations that at a given time are experiencing the horizon re-entry contributes to the tidal field, scaling as $k^2$ for $k\rightarrow 0$. Now that we have the contribution of tensor modes to the local tidal field, we need to make assumptions on how the galaxy shape field responds to a change in this tidal field. 
\paragraph{The instantaneous response.} One basic approach is to extend the linear alignment model to tensor modes and assume that the response to a change in $t_{ij}$ is  ``instantenous'', and therefore the projected, density-weighted galaxy shape field can be written as
\begin{equation}
    \tilde \gamma_{ij}(\hat \bn, z_O) =  \mathcal P_i^\ell \,\mathcal P_j^m \,\Bigr[\, b_K\, K_{ij}(\hat \bn,z_O)(1 + b_\delta\, \delta(\hat \bn,z_O) ) + b_t\, t_{ij}(\hat\bn,z_O)(1 + b_\delta\, \delta(\hat \bn,z_O) )\, \Bigr] \, ,
\end{equation}
where we have dropped the $\delta^2$ and $K^2$ from the galaxy bias expansion, Eq. \eqref{eq:deltagexp}, as we have seen from the previous Section  \ref{sec:scalar} that at this order they only renormalize bias parameters.
It is important to notice here that, because the time evolution of the transfer function $T_h$ depends on the long-wavelength mode $k_L$ (cf. Eq. \eqref{eq:ttransfer}), the amplitude of the response of $\tilde \gamma_{ij}$ to changes in the primordial tensor perturbation $h^{(0)}_{ij}$ depends itself on $k$. In other words, we cannot factorize the time component in Fourier space in corresponding growth factors for the evolution of tensor perturbations, as it is usually done for scalar perturbations.  

\paragraph{The fossil effect.} Another approach was considered by \cite{Schmidt:2013nsa}, where they calculated the effect of such a tidal tensor on the second order density field $\delta$ to be
\begin{equation}
    \delta^{(2)}_t(\bx,\eta) = h_{ij}^{(0)}(\vx)\left[ \alpha(k_L,\eta)\,\frac{\partial^i\partial^j}{\nabla^2} + \beta(k_L,\eta)\, x^i\, \partial^i\,\right] \delta^{(1)}_s(\bx,\eta) \, ,
\end{equation}
where $\delta^{(N)}_X$ indicates the matter density field at order $N$ for $X=s,t$ scalar and tensor perturbations, respectively and $\alpha$ and $\beta$ are given, assuming matter domination, by
\begin{align}
    \alpha(k_L,\eta) &= \frac 25 + 18\, \frac{\cos(k_L\eta)}{(k_L\eta)^4} + 6 \frac{\sin(k_L\eta)}{(k_L\eta)^3}\left[1-\frac{3}{(k_L\eta)^2}\right],\\
    \beta(k_L,\eta) &= \frac 12 + \frac32\, \frac{\cos(k_L\eta)}{(k_L\eta)^2} + \frac32 \frac{\sin(k_L\eta)}{(k_L\eta)^3} \, .
\end{align}
These functions reflect the fact that the tidal field sourced by tensor perturbations depends on a time integration over the past history of the tensor mode, rather then on the ``instantaneous'' value of $t_{ij}$. For more details on the derivation of $\alpha$ and $\beta$, we refer to the original calculation \cite{Schmidt:2013gwa}.
%, here instead we summarize their main physical features: 
%\begin{itemize}
%    \item First of all, they have the expected behaviour of going to zero on superhorizon scales, as required by the equivalence principle.
%    \item Their contribution peaks at horizon crossing, when the mode enters the horizon, but differently than in the instantaneous case of Eq. \eqref{tidal_tens} it does not decay for $k_L \eta \gg 1$, rather they asymptote to constant values, $\alpha_\infty=2/5$ and $\beta_\infty=1/2$ in matter domination. For this reason, their effect on small-scale matter perturbations has been called ``fossil effect'' in \cite{Masui:2010cz,Jeong:2012df}.
%    \item The term proportional to $\alpha$ is the proper effect of the tidal field on small-scale fluctuations. On the other hand, the term proportional to $\beta$ is a displacement term: the tidal field slightly changes the mapping from Lagrangian to Eulerian positions.
%    \item Similarly to the case of instantaneous alignment, the transfer functions $\alpha$ and $\beta$ depend non trivially on the long-wavelength mode $k_L$, such that the time dependence cannot be factorized from it. The factorization is however possible for modes well within the horizon, $k_L\eta\gg 1$, since in this regime the two functions go to a constant.
%\end{itemize}
If we take seriously the fossil effect of tensor modes on small-scale matter perturbations, we should believe that galaxy shapes respond similarly to these perturbations. Indeed,  the calculation in FNC frame of the effect of long-wavelength perturbation $k_L$ on small-scale density perturbations is equivalent, at least in procedure, for scalar and tensor perturbations. 
%For scalar perturbations, the trace-free contribution from $K_{ij}$ to the second order matter density field reads
%\begin{equation}\label{eq:f2shear}
%    \delta^{(2)}_s(\bx,\eta) = \frac 27 \left(\frac{\partial_i\partial_j}{\nabla^2} \delta^{(1)}_L(\bx,\eta)\right)\frac{\partial_i\partial_j}{\nabla^2} \delta^{(1)}_s(\bx,\eta) \, .
%\end{equation}
%Assuming a LA model, we know how scalar perturbations source an external tidal field which affects galaxy shapes, i.e.
%\begin{equation}\label{eq:iascalar}
%    \gamma^s_{ij}(\vx, \eta) = b_s \, \mathcal P_i^\ell \,\mathcal P_j^m \left(\frac{\partial_\ell\partial_m}{\nabla^2}\delta(\vx,\eta)\right)
%\end{equation}
%for some response $b_s$, which in our LA model we have called $b_K$. 
Consequently, in \cite{Schmidt:2013gwa} it is argued that the way in which galaxy shapes are affected by tensor perturbations should be matched to the corresponding response calculated in the case of scalar perturbations, therefore obtaining
%\footnote{ In \cite{Schmidt:2013gwa} it is assumed that the response factors are equivalent, $b_h \equiv b_s$, and they also match for the factor of $2/7$ in Eq. \eqref{eq:f2shear}. Here we keep a more agnostic approach, considering $b_h$ and $b_s$ as separate parameters.}
\begin{equation}\label{eq:iatensor}
    \gamma^t_{ij}(\vx, \eta) = b_h \, \alpha(k_L,\eta)\, \mathcal P_i^\ell \,\mathcal P_j^m\, h_{\ell m}^{(0)}(\vx) \, .
\end{equation}
If we apply this ansatz, the full expansion in the projected density weighted galaxy shape field is
\begin{equation}\label{eq:fulltens}
        \tilde \gamma_{ij}(\hat \bn, z_O) =  \mathcal P_i^\ell \,\mathcal P_j^m \,\Bigr[\, b_K\, K_{ij}(\hat \bn,z_O)(1 + b_\delta\, \delta(\hat \bn,z_O) ) + b_h\, \alpha(k_L, z_O)\, h^{(0)}_{ij}(\hat \bn,z_O)(1 + b_\delta\, \delta(\hat \bn,z_O) )\, \Bigr] \, .
\end{equation}
We employ this second prescription for our computations, calculating the numerical expression for $\alpha(k_L, z_O)$ including radiation and $\Lambda$, i.e. without assuming matter domination, as shown in \cite{Schmidt:2013gwa}. Having the response of $\tilde \gamma_{ij}$ to primordial tensor modes, we can go on with a similar procedure as for the previous section by applying spin-lowering operators to get that the leading order contribution to the galaxy shape field from tensors in Fourier space is 
%Consistently to how we chose to expand the galaxy shape field, we do not take this route and add directly the tensor tidal field $t_{ij}$ to the expansion \eqref{eq:exptilde}. Following the same procedure as for the late-time induced terms, we can calculate the leading order contribution sourced by tensors in the projected scalar galaxy shape correlation to be
\begin{equation}
    \tilde\gamma(\vnhat,\vk) \supset \frac 14 b_h\, \alpha(k)\, \sum_{p=-1,1} h^{(0)}_{2p}(\bk)\, Q_{2 p}(x)\,e^{ix\mu}\,(1-\mu^2)\,e^{i2p\phi} \, ,
\end{equation}
where we will drop the dependence on redshift in $\alpha$ from now on and $Q_{\pm 2}(x)$ are defined in Appendix \ref{app:scalars}, and $h^{(0)}_{\pm 2}(\bk) = h_{R/L}(\bk)$. We now have all the ingredients to write all the contributions from tensor modes to the galaxy shape power spectrum. At order $\mathcal O(P_X^2)$, where $X=\delta,h$, and respecting the LA model, we have two terms.
\paragraph{The $\langle h h \rangle$ term.} At linear order on $P_h$ we find
\begin{align}\label{eq:bbpower}
    C_\ell^{EE,(hh)} &= \frac{b_h^2}{16}\, \int \frac{dk}{2\pi}\, k^2\, \alpha^2(k)\,P_h(k) |F^{E2}_\ell(x)|^2\\
    C_\ell^{BB,(hh)} &= \frac{b_h^2}{16}\,\int \frac{dk}{2\pi}\,k^2\, \alpha^2(k)\, P_h(k) |F^{B2}_\ell(x)|^2 \, ,
\end{align}
where we used $P_{\pm 2}(k) = P_h(k)/8$ and $F^{X2}(k)$ are the same as the ones used for the scalar-induced correlations in the previous Section \ref{sec:scalar}.
\paragraph{The $\langle h \delta h \delta \rangle$ term.} Similarly to the case of $K_{ij}$, this term involves the convolution of the primordial tensor field $h$ with the galaxy density field $\delta_g$,
\begin{equation}
    \int_{\vq} P_h(q)P_\delta(|\vk-\vq|)\, ,
\end{equation}
and similarly to before for $k\rightarrow 0$ a shot-noise contribution arises. The correlators for this term read
\begin{align}\label{eq:hdhdE}
    C_\ell^{EE,(\delta h)^2} &= b_{h}^2\, b_\delta^2  \int \frac{dk}{2\pi}\, k^2 \left[ \frac{(\ell-2)!}{(\ell+2)!} |F^{E0}_\ell(x)|^2 T_0(k) + \frac{1}{\ell^2+\ell-2}\, |F^{E1}_\ell(x)|^2 T_1(k)+  |F^{E2}_\ell(x)|^2 T_2(k)\right]\\
    C_\ell^{BB,(\delta h)^2} &= b_{K}^2\, b_\delta^2  \int\frac{dk}{2\pi}\, k^2\,\left[  \frac{1}{\ell^2+\ell-2}\, |F^{B1}_\ell(x)|^2 T_1(k)+  |F^{B2}_\ell(x)|^2 T_2(k)\right]\, ,\label{eq:hdhdB}
\end{align}
where
\begin{align}
    T_n(k) & = \frac{k^3}{(2\pi)^2}\int_0^\infty dr \int_{-1}^{1}d\mu\, \alpha^2(r k)\, P_h(r k) P_\delta( k\sqrt{1+r^2-2 r \mu}) \tilde T_n(r,\mu)\, ,
\end{align}
and we define the kernels $\tilde T_n$ in Appendix \ref{app:scalars}. We have verified, even before subtraction of the shot-noise, that this term gives a negligible contribution to the total power spectrum, we therefore neglect it.

\begin{figure}[ tbp]
	\centering
	\includegraphics[width=0.495\textwidth]{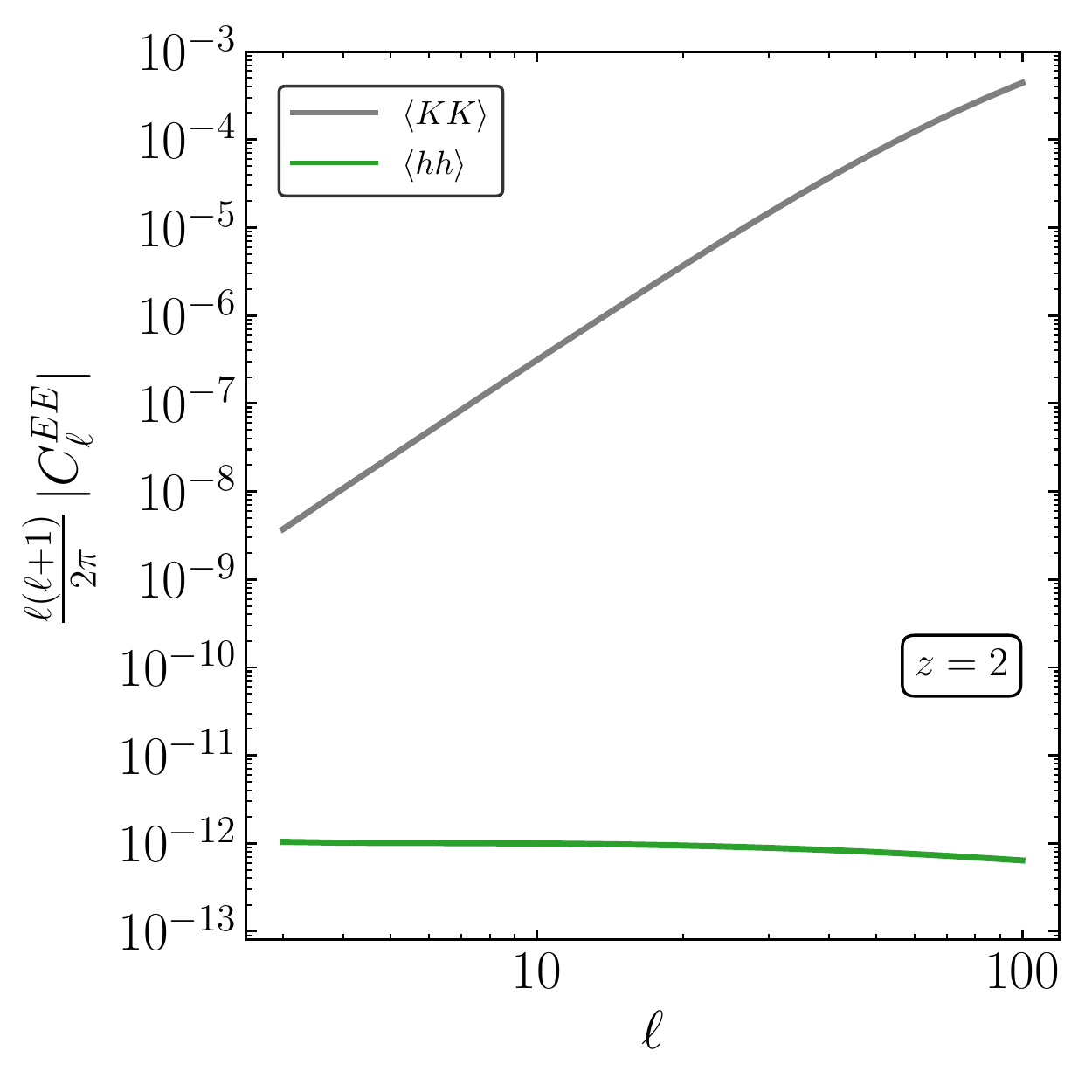}
	\includegraphics[width=0.495\textwidth]{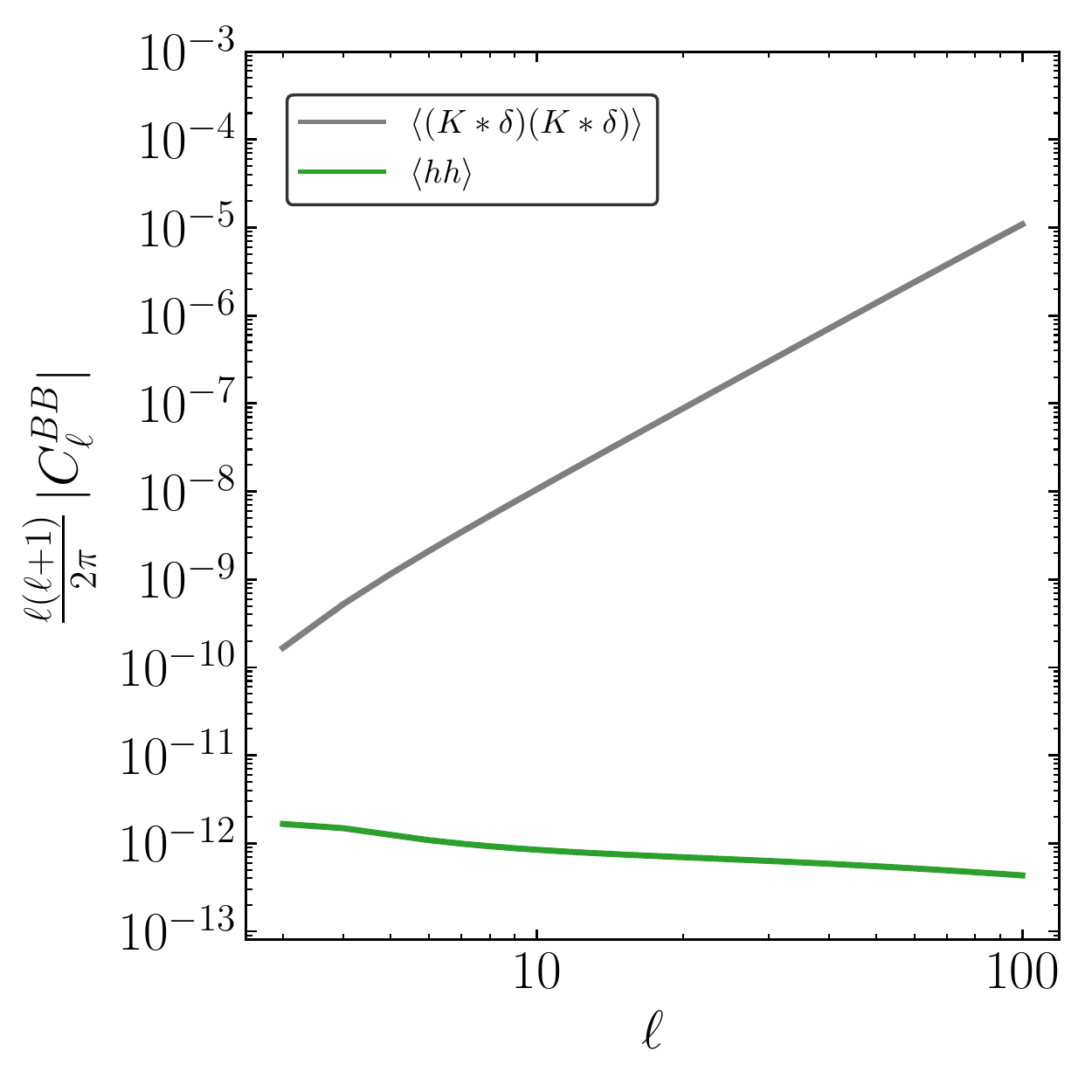}
	\caption{EE (left) and BB (right) density-weighted galaxy shape power spectrum at two different redshifts as generated in the LA model by the tidal shear field $K_{ij}$ and by the tensor tidal field $t_{ij}$ using the prescription of Eq. \eqref{eq:iatensor}.}
	\label{fig:tensorvsscalar}
\end{figure}

We show the EE and BB galaxy shape power spectrum as sourced by tensor (Eq. \eqref{eq:bbpower}) and scalar (Eq. \eqref{eq:bbscalars}) perturbations in Figure \ref{fig:tensorvsscalar}, where we choose all bias parameters to be unity, $b_K = b_\delta = b_h = 1$.\footnote{ While this is not a realistic scenario, there is no reason at this stage to expect that $b_K$ is much greater than $b_h$ and anyway their ratio would be ultimately only constrained by data. As for $b_\delta$, changes of order unity are expected, but would not affect qualitatively our results.}
%We also show the estimated $1\sigma$ uncertainty following Eq. \eqref{uncertainty_signal}. 
Although different in shape, the contribution from tensor modes is subdominant with respect to the one induced by $K_{ij}$ even in the case of the BB power spectrum, which is induced by $K_{ij}$ at order $\mathcal O(P_\delta^2)$. %The primordial contribution is also somewhat lower than the observability threshold we consider. 
For this reason, we are not showing higher order terms involving $\mathcal O(P_\delta \times P_h)$, which are negligibly small. We would also like to stress that we are comparing strictly intrisic correlations in galaxy shapes. More correlations among shapes arise when considering also the lensing contributions, as shown e.g. in \cite{Schmidt:2012ne}. Figure \ref{fig:tensorvsscalar} shows that even before accounting for lensing contaminations to the primordial signature, there are intrinsic ones that are expected to be larger in amplitude, though different in shape.

\section{Chiral Gravitational Waves and intrinsic alignments}\label{sec:chiral}

In Section  \ref{sec:scalar}, we have shown that $B$ modes arise at order $P^2$ in the density-weighted galaxy shape power spectrum as a consequence of the convolution of the tidal shear field $K_{ij}$ with the galaxy density field $\delta_g$. It is important to specify that this is not a specific feature of our approximation of considering the LA model and computing density-weighted galaxy shape statistics: the convolution would appear generically in an EFT expansion, as already shown in \cite{Vlah:2019byq}. These scalar-induced B modes in the galaxy shape power spectrum are a contaminant to the primordial signature coming from tensors, as shown in Figure \ref{fig:tensorvsscalar}. 
%Other contaminants have been identified in the projected component in earlier studies \cit. 
This finding motivates a search of a setup where the primordial signature is the only, or at least dominant, source. In this section, we consider parity-breaking primordial gravitational waves as a source of intrinsic alignments of galaxy shapes. We will show that E-B intrinsic correlations are generated by these parity-breaking models. 

\paragraph{Chirality from inflation.}
Gravitational waves produced in an inflationary context are predicted to be unpolarized by the standard slow-roll models of inflation (see e.g. the review \cite{Guzzetti:2016mkm}), where parity is a symmetry of the theory. However, we currently lack strong observational constraints on the level of chirality of primordial tensor modes. In fact, we have just forecasts about testing chirality of primordial gravitational waves with CMB data (see e.g. \cite{Gerbino:2016mqb, Bartolo:2018elp}) and interferometers (see e.g. \cite{Crowder:2012ik, Smith:2016jqs, Thorne:2017jft}). 

Chirality in the context of gravitational waves can be defined at linear level as the relative difference between the R and L-handed primordial tensor power spectra
\begin{equation}\label{eq:chi}
\chi(k) = \frac{P_R(k)-P_L(k)}{P_R(k)+P_L(k)} \, ,
\end{equation}
being
\begin{equation}\label{eq:prl}
    P_{R/L}(k) = \langle h_{R/L}(k) h_{R/L}^*(k) \rangle \, .
\end{equation}
We expand on model building efforts for various predictions of $\chi \neq 0$ in Appendix \ref{app:chiralmodels}, but our results are generic as long as the inflationary model produces chirality of the type of Eq. \eqref{eq:chi}\footnote{While a maximum chirality $\chi=1$ is generically hard to achieve for inflationary models where the tensor perturbations are not sourced by a spectator field, it has been shown for instance in \cite{Maleknejad:2018nxz} and references therein that primordial gauge fields, such as an axion-SU(2) gauge field, can enhance the parity breaking during inflation, raising $\chi$ to unity.}.

\subsection{Chiral gravitational waves in the galaxy shape power spectrum}
\begin{figure}[ tbp]
	\centering
	\includegraphics[width=0.495\textwidth]{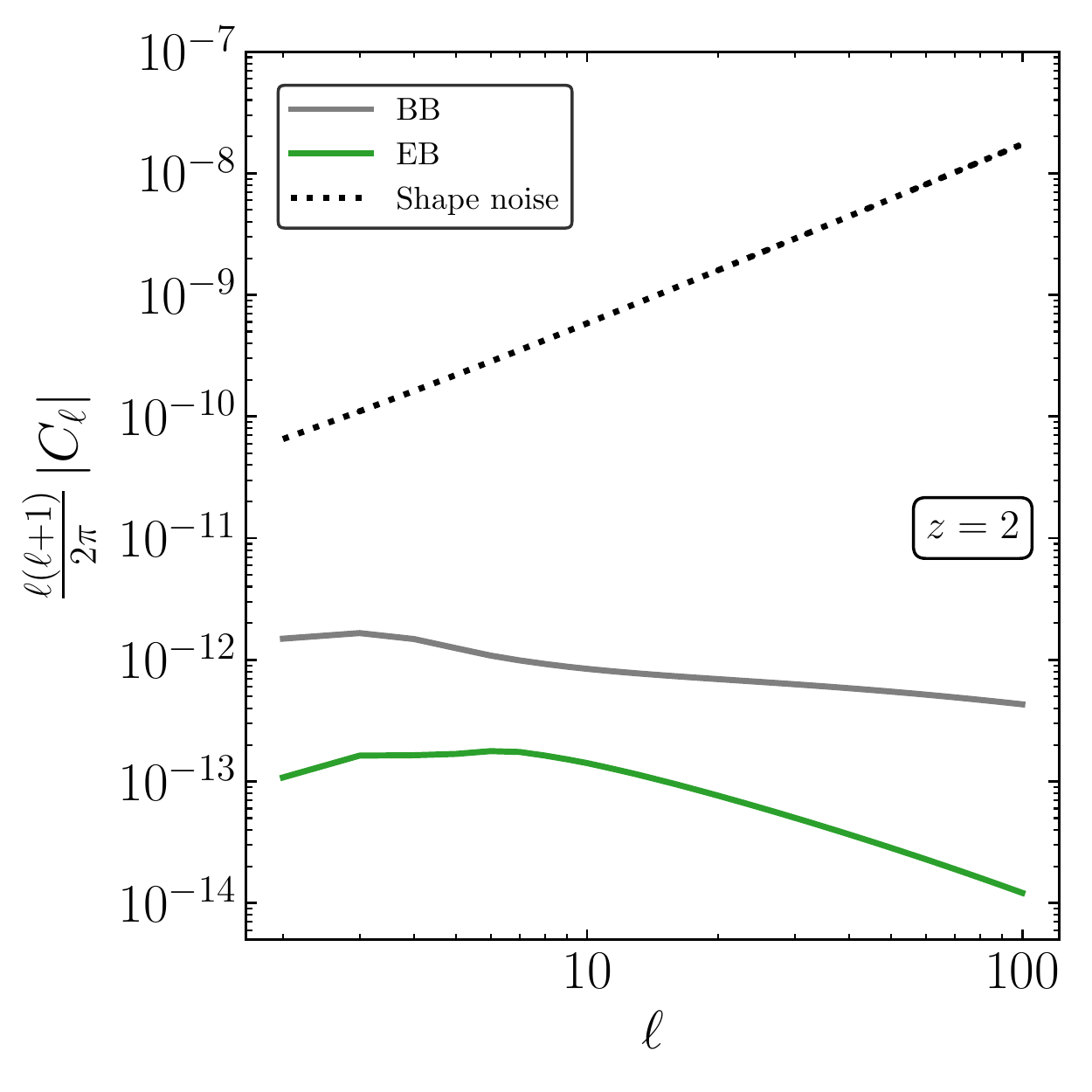}
	\includegraphics[width=0.495\textwidth]{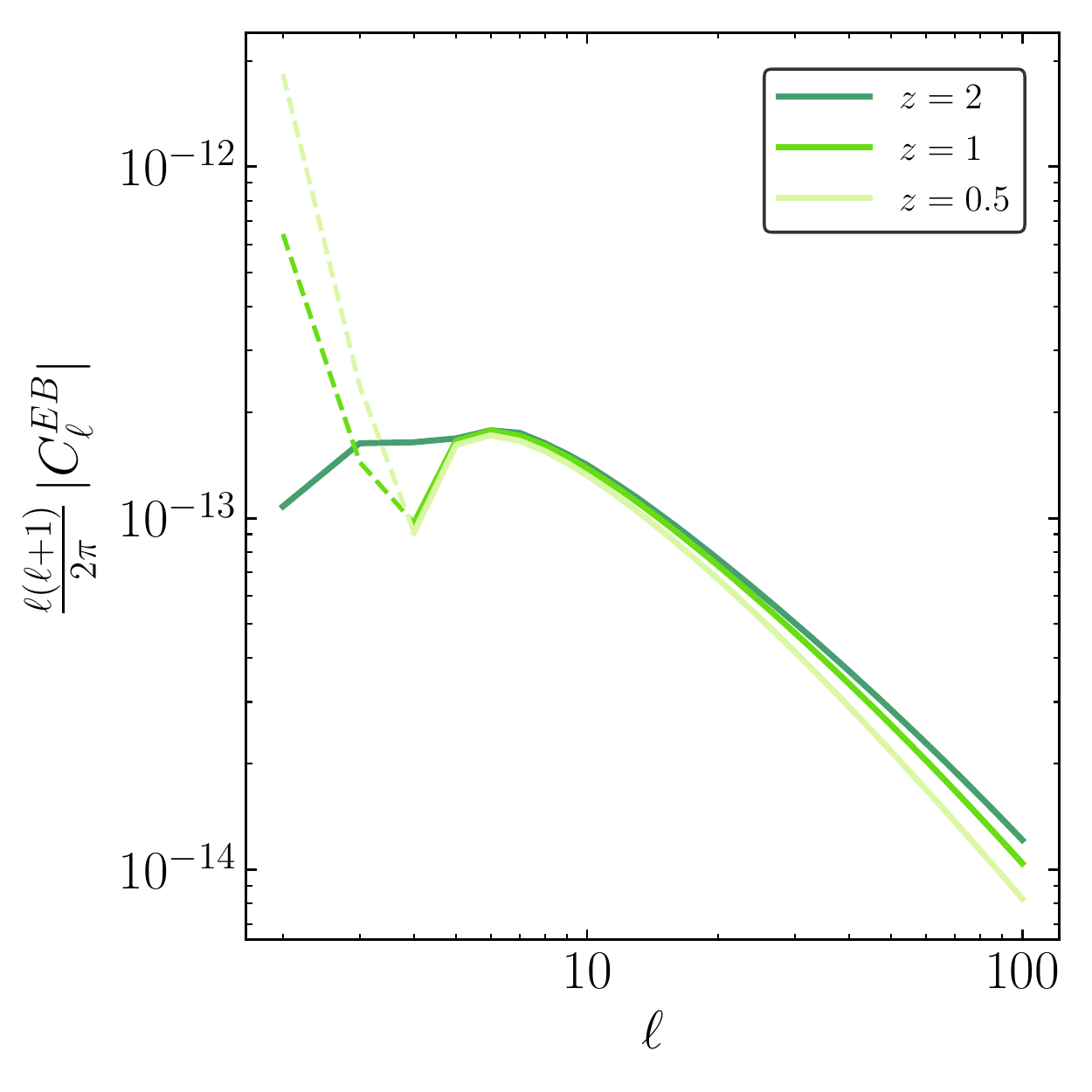}
	\caption{Left Panel: BB and EB galaxy shape power spectra at redshift $z=2$, where the parity-breaking is that of Eq. \eqref{eq:chi} with maximum chirality, $\chi=1$. Right Panel: Comparison of EB power spectra at three different redshifts for the same model. Dashed lines indicate negative values.}
	\label{fig:chiral}
\end{figure}

First, let us notice that correlations of $K_{ij}$, or of any other combination of $K_{ij}$ and $\delta$ producing only matter density power spectra $P_\delta$ can not give a parity-breaking contribution to the galaxy shape power spectrum, just because by definition  $\delta(\mathbf x)$ evolving in a FLRW Universe is a real scalar field and thus invariant under parity\footnote{We can see this concretely by observing how the harmonic coefficients $a_{\ell m}$ and the derivative functions $Q_n$ transform under parity\cite{Schmidt:2012ne}. For $\hat n\rightarrow -\hat n$ we have
\begin{align}
    a^{E}_{\ell m} &\rightarrow (-1)^\ell a^{E}_{\ell -m} \,,\\
    a^{B}_{\ell m} &\rightarrow -(-1)^\ell a^{B}_{\ell -m} \, ,\\
\end{align}
so that
\begin{equation}
    \langle a^{E}_{\ell m} a^{B}_{\ell m} \rangle \rightarrow -\langle a^{E}_{\ell -m} a^{B}_{\ell -m} \rangle \, .
\end{equation}
At this point, assuming the parity symmetry we get
\begin{equation}
    C_{\ell}^{E B} = \sum_m \langle a^{E}_{\ell m} a^{B}_{\ell m} \rangle \stackrel{\hat n\rightarrow -\hat n}{=} - \sum_m \langle a^{E}_{\ell -m} a^{B}_{\ell -m} \rangle =  -  C_{\ell}^{E B} = 0\, .
\end{equation}}. We therefore do not expect to see a finite correlation between $E$ and $B$ modes in the galaxy shape power spectrum, unless a primordial process, such as the one presented above, violates parity at the level of tensor perturbations\footnote{Observations of SDSS galaxies at low redshift seem to hint at a potential parity violation of the observed spin of spiral galaxies \cite{Longo:2011nk,Shamir:2012ej,Shamir:2019cdy,Shamir:2019fmw}. However, the strength of the violation and the low-redshift and small scales at which it is observed suggests that it is not of primordial origin.}.  
Given the parity-breaking primordial tensor power spectrum from Eq. \eqref{eq:prl}, we can compute the cross-correlation of E and B modes. The only difference with respect to the EE and BB power spectra sourced by primordial gravitational waves is that the EB correlation depends on the difference $P_R-P_L = \chi\, P_h /8$. We therefore find
\begin{equation}
    C_\ell^{EB} = \chi\,\frac{b_h^2}{16} \, \int \frac{dk}{2\pi}\, k^2\, \alpha^2(k)\,P_h(k)\, F^{E2}_\ell(x)\, F^{B2}_\ell(x)\, ,
\end{equation}
where we now neglect the term proportional to $\P_h \times P_\delta$ as we have already determined that it is small. We show a comparison of the BB and EB power spectra sourced by tensor perturbations only in Figure \ref{fig:chiral}, where as before we choose $b_h=1$. Since we take the maximal amount of chirality for these plots, $\chi=1$, the difference in shape and amplitude is entirely given by the different combinations of transfer functions $F^{E2}_\ell$ and $F^{B2}_\ell$.
In order to determine whether these signatures are, at least in principle, observable, we can argue that they should be at least of the same order, or higher, as the shot-noise usually computed for imaging surveys measuring the intrinsic ellipticity of galaxies, the so-called  ``shape noise'' (see for instance \cite{Chang:2013xja}). Indeed, until now, we have worked on large-scales, ignoring the fact that stochasticity is produced by small-scale perturbations and affects the formation of galaxies, and therefore their intrinsic ellipticity. These stochastic contributions can be systematically accounted for order by order in effective descriptions in a similar way as done for the galaxy bias, as explained in \cite{Vlah:2019byq}. In imaging surveys, the leading, scale independent, contribution to the galaxy shape power spectrum is usually expressed as
\begin{equation}
\sigma_{\gamma}^2 = \frac{\sigma_e^2}{\bar n} \, ,    
\end{equation}
where, for a given fraction of sky considered, $\sigma_e$ is the RMS (Root Mean Square) intrinsic ellipticity of galaxies, and $\bar n$ is the number of source galaxies per steradian. In the approximation in which the measured signal is dominated by this shot-noise, and neglecting systematics from the instrument itself, the $1\sigma$ uncertainty on the measured signal is
\begin{equation} \label{uncertainty_signal_approx}
\Delta C^{E/B}_{\ell, \rm measured} \simeq \sqrt{\frac{2}{(2 \ell + 1) f_{\rm sky}}}  \sigma_{\gamma}^2 \, .
\end{equation}
For our estimations, we consider an LSST-like survey with $\sigma_e^2=0.26$, $f_{\rm sky}=0.36$ and $\bar n= 31$ galaxies/arcmin$\mbox{}^2$ and median redshift $z=0.93$ \cite{Chang:2013xja}. Since we do not account for the redshift distribution $dN/dz$ of the survey, we just compute power spectra at $z=1$ which is close to the median redshift. We therefore use this setup just as an approximate threshold of observability of our signatures, determining that the signature is $1-2$ order of magnitudes below the threshold\footnote{These prospects might be improved upon cross-correlating the galaxy shape field with the CMB polarization field, as studied for the case of EE and BB correlations in \cite{Chisari:2014xia}. Their work show, however, that the CMB auto-correlation contains most of the constraining information and we do not expect to find significantly different results for the EB correlation.}.
An important question to raise is whether we still have the shape-noise signal for the EB power spectra. Following the argument above, one might think that, without any parity breaking processes arising during galaxy formation, there should not be any stochasticity in the EB correlation, thus causing the shape noise signal of the EB channel to be vanishing. This, in principle, reduces the unavoidable uncertainty \eqref{uncertainty_signal_approx} with respect to the EE and BB channels.

\section{Tensor non-Gaussianities and intrinsic alignments}\label{Imprint_no_Gaus}

Primordial tensor non-Gaussianities have been object of several studies and provide an interesting window to interactions taking place during inflation at very high energies \cite{Maldacena:2002vr,Maldacena:2011nz,Soda:2011am,Gao:2011vs,Senatore:2011sp,Cook:2013xea,Camanho:2014apa,Namba:2015gja,Akita:2015mho,Guzzetti:2016mkm,Bartolo:2017szm,Agrawal:2017awz,Dimastrogiovanni:2018gkl,Biagetti:2017viz,Franciolini:2017ktv,Baumann:2017jvh,Bordin:2018pca,Goon:2018fyu,Ozsoy:2019slf,Fujita:2019tov,Baumann:2019oyu}. The galaxy shape field bispectrum $\langle g\, g\, g \rangle$ or cross-correlations of shape and density such as $\langle g\, g\, \delta \rangle$ and $\langle g\, \delta\, \delta \rangle$ contain the direct information on tensor non-Gaussianities $\langle h\,h\,h\rangle$ and the respective cross-correlations with the scalar perturbation $\langle h\, h\, \zeta\rangle$ and $\langle h\,\zeta\,\zeta\rangle$ through the modelling of how $g_{ij}$ responds to tensor perturbations we explained in Section  \ref{sec:tensors}. While we defer investigation of the galaxy shape bispectrum to future work, in this section we show that the galaxy shape power spectrum is itself  sensitive to tensor non-Gaussianities of the type $\langle h\,h\,\zeta\rangle$ and $\langle h\,\zeta\,\zeta\rangle$ due to the density-weighting of the galaxy shape field. We calculate their imprint on intrinsic alignments for two models where these correlations have enhanced amplitude during inflation.

\subsection{Tensor non-Gaussianity from inflation}
In Section  \ref{sec:tensors}, we wrote down the expansion of the projected density-weighted galaxy shape field, Eq. \eqref{eq:fulltens}, in the LA model. Similarly to Eq. \eqref{eq:intuit}, we can momentarily drop indices and schematically write down  all the possible terms sourced by tensors contributing to the power spectrum as
\begin{align}\label{eq:expintu}
    \langle \tilde \gamma\, \tilde \gamma \rangle \supset &\, b_h^2\, \alpha^2\, \langle h\, h \rangle + 2 \,b_K\, b_h\,  \alpha\, \langle K\, h\, \rangle  + 2\, b_K\, b_h\,  \alpha\,  b_\delta\, \bigr[\langle (K * \delta)\, h \rangle +\langle K \, (h*\delta) \rangle\bigr] \nonumber\\
    &  + 2\, b_h^2\, \alpha\, b_\delta\,\langle h\, (h*\delta) \rangle  + 2\, b_K\, b_h \, \alpha^2\, b_\delta^2 \langle (K*\delta)\, (h*\delta) \rangle + b_h^2\,  \alpha^2\, b_\delta^2\, \langle (h*\delta)\,(h*\delta) \rangle \, ,
\end{align}
where the first term is the leading contribution which we already discussed in Sections  \ref{sec:tensors} and  \ref{sec:chiral}. The second term, $\langle K\, h \rangle$, is non-zero only in particular anisotropic primordial setups where the scalar-tensor cross correlator $\langle \zeta h \rangle$ is sourced. Example of models predicting it are e.g. \cite{Ohashi:2013qba,Bartolo:2015dga,Almeida:2019hhx}. However, in these models, an enhanced amplitude of $\langle \zeta h \rangle$ is associated with large anisotropies in the scalar cross-correlator $\langle \zeta \zeta \rangle$ and in primordial non-Gaussianities, and observations on the statistically anisotropic modulations of the CMB (see e.g. \cite{Akrami:2018odb}) have put tight constraints on these anisotropies. The third and fourth term in \eqref{eq:expintu} are sourced by primordial non-Gaussian correlators of the type  $\langle h\, h\, \zeta\rangle$ and $\langle h\,\zeta\,\zeta\rangle$, while the last two terms are sourced by inflationary trispectra. The last term is also sourced by the product of the tensor and scalar power spectrum. 

\paragraph{Parity-Breaking Tensor non-Gaussianities.} In Section  \ref{sec:tensors} we argued that EE and BB primordial power spectra are subdominant with respect to their counterparts sourced by scalar perturbations through $K_{ij}$. In the same spirit of looking for a distinctive parity breaking signature in the power spectrum, we look at parity-breaking bispectra. In this case, we specialize to a model generating parity breaking signatures via the Chern-Simons modified gravity term \eqref{Chern-Simons}, which was developed in \cite{Bartolo:2017szm}.
Within this model, parity breaking is generated in the power spectrum in the form of Eq. \eqref{eq:chi}, where in this case $\chi$ is a function of the Chern-Simons mass, and in the bispectrum statistics, providing a source for the parity breaking term $\langle h\, h\, \delta \rangle$. 
In Ref. \cite{Bartolo:2017szm}, the exact shape function of the parity-breaking contribution to the primordial tensor-tensor-scalar bispectrum statistics was computed to be
\begin{align} 
\langle h_{R/L}(\vec{k}_1) h_{R/L}(\vec{k}_2) \zeta(\vec{k}_3) \rangle
&= (2\pi)^3 \delta^{(3)} \, \left(\vec k_1 + \vec k_2 + \vec k_3 \right) B^{R/L}_{hh\zeta}(k_1, k_2, k_3) \, ,\\
\langle h_{R/L}(\vec{k}_1) h_{L/R}(\vec{k}_2) \zeta(\vec{k}_3) \rangle &= 0 \,, 
\end{align}
where 
\begin{equation} \label{eq:hhzeta}
B^{R/L}_{hh\zeta}(k_1, k_2, k_3) = \mp  
\frac{25 \pi^4}{768} {\cal A}_s^2 \left( r^2 \Pi \right)
\frac{(k_1 + k_2)}{k_1^2 k_2^2 k_3^3} \frac{\cos \theta (1 - \cos \theta)^2}{2} \, ,   
\end{equation}
where 
\begin{equation}
\cos \theta = \frac{k_3^2 - k_2^2 - k_1^2}{2 k_1 k_2} 
\end{equation}
is the cosine of the angle between the momenta $\vec k_1$ and $\vec k_2$ forming a triangle configuration with $\vec k_3$ and $\Pi$ is a dimensionless parameter defined as
\begin{equation} \label{Pi}
\Pi = \frac{96 \pi}{25} H^2 \frac{\partial^2 f(\phi)}{\partial^2 \phi}\, ,    
\end{equation}
being $f(\phi)$ the coupling function in Eq. \eqref{Chern-Simons} and $\partial^2 f(\phi)/\partial^2 \phi$ its second order derivative. A priori, the quantity $\partial^2 f(\phi)/\partial \phi^2$ can be scale dependent, but in this work we will assume it to be scale independent for simplicity. In order to maintain perturbativiy of the theory generating these interactions (see Appendix \ref{bound:Pi} for more details), the amplitude of this non-Gaussianity is theoretically bounded as
\begin{equation}
\Pi \lesssim \left(\frac{0.1}{r}\right) \times 10^6 \, .    
\end{equation}
Moreover, the expression of the bispectrum \eqref{eq:hhzeta} has to be corrected in the so-called squeezed limit where the momentum of the scalar perturbation $\zeta$ is much smaller than the momenta of the two gravitons (i.e. $k_3 \ll k_1 \simeq k_2$). In fact, it is well known from the literature (see e.g.  \cite{Tanaka:2011aj,Kehagias:2013yd,Peloso:2013zw,Dai:2013kra,Creminelli:2013mca,Creminelli:2013cga,Pajer:2013ana,Hinterbichler:2013dpa,Mirbabayi:2014zpa,Dai:2015rda,Dai:2015jaa,Bordin:2016ruc,Cabass:2016cgp,Cabass:2018roz}) that in the squeezed limit the leading order value of the primordial bispectra can be reabsorbed leaving only small physical contributions of order $(k_L/k_S)^2$ (see  Appendix \ref{squeezed_limit} for an estimate of this correction for our model).

\paragraph{Squeezed Tensor non-Gaussianities.} Similarly to the case of primordial non-Gaussianity sourced by scalar perturbations, consistency relations constrain the amplitude of tensor non-Gaussianity in the squeezed limit to be small \cite{Maldacena:2002vr,Pajer:2013ana}. The breaking of the consistency relations for scalar non-Gaussianities is usually related to multi-field models of inflation, the most popular example being local-type primordial non-Gaussianity \cite{Salopek:1990jq}. For tensor non-Gaussianities, it is somewhat harder to break the relative consistency relations,  as they are violated only when adiabaticity is broken by light tensor perturbations \cite{Bordin:2016ruc}. Therefore, multiple scalar fields do not help \cite{Dimastrogiovanni:2015pla} and the Higuchi bound  forbids the existence of light spin-2 fields in De Sitter (DS) \cite{Higuchi:1986py}. One way to violate the consistency relations is hence to break the DS isometries \cite{Bordin:2018pca} and therefore allow for light particles with spin during inflation. Another way is through partially massless higher-spin particles \cite{Lee:2016vti,Kehagias:2017cym, Franciolini:2017ktv, Goon:2018fyu}. Assuming that one of these scenarios take place, the form of the squeezed tensor-scalar-scalar bispectrum is
\begin{equation}\label{eq:hzetazeta}
    B_{h \zeta\zeta}(k_1,k_2,k_3)\Bigr|_{k_1 \ll k_2,k_3} = f_{\rm NL}^h\, \left(\frac{k_1}{k_2}\right)^{\frac 32-\nu}\, P_h(k_1) \, P_\zeta(k_2)\, ,
\end{equation}
where $\nu = \sqrt{9/4-(m/H)^2}$, being $m$ the mass of the particle exchanged in the process and $H$ the Hubble radius during inflation. Being $k_1/k_2 \ll 1$, the maximum amplitude is reached for massless particles, for which $\nu\rightarrow 3/2$ and the ratio of the long mode over the short one vanishes.
%As we will see starting from the next subsection, primordial bispectrum \eqref{eq:hhzeta} gives an additional parity breaking contribution to the intrinsic shear power spectrum, while no contributions are made by the scalar-scalar-tensor and tensor-tensor-tensor primordial statistics.

\subsection{Tensor non-Gaussianities in the galaxy shape power spectrum}
\begin{figure}[ tbp]
	\centering
	\includegraphics[width=0.495\textwidth]{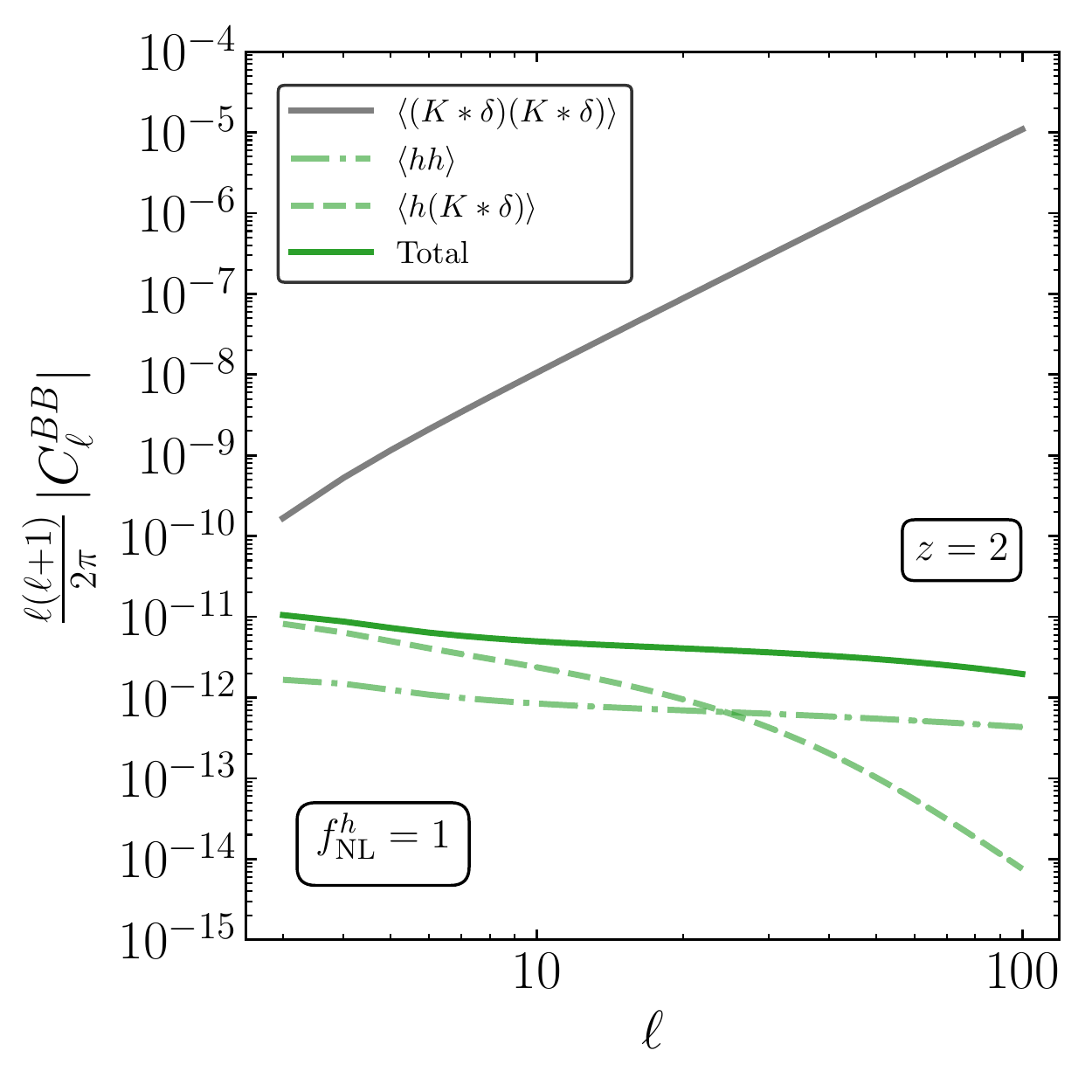}
	\includegraphics[width=0.495\textwidth]{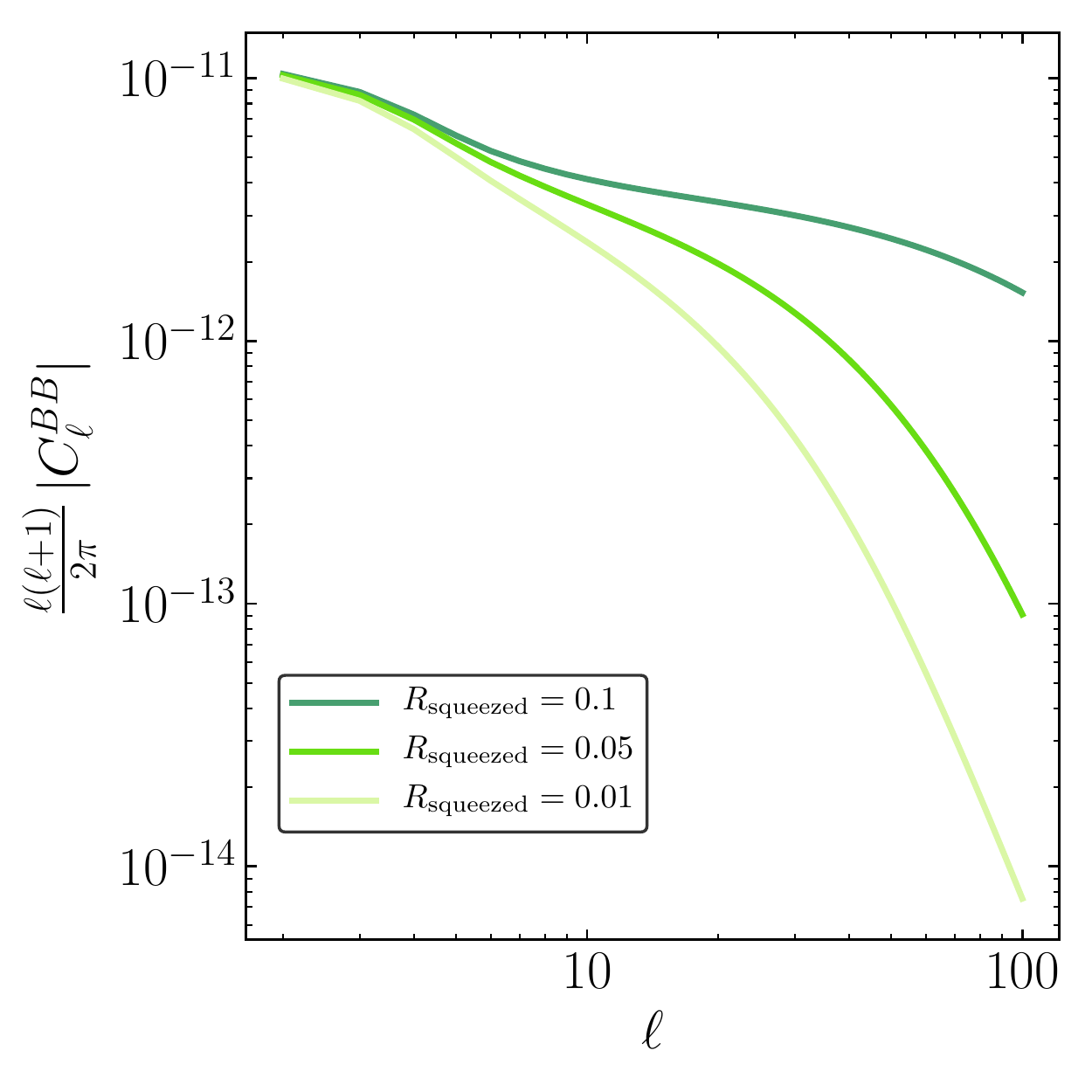}
	\caption{Left Panel: The contributions from the primordial  bispectrum of Eq. \eqref{eq:hzetazeta} with $f^h_{\rm NL}=1$ (green dashed), from the tensor power spectrum (green dotted-dashed) and their sum (green solid) compared with the scalar-induced BB power spectrum (gray solid) at redshift $z=2$. Right Panel: Contribution from the primordial  bispectrum of Eq. \eqref{eq:hzetazeta} with $f^h_{\rm NL}=1$ to the galaxy shape power spectrum for different values of $R_{\rm squeezed}$ at redshift $z=2$.}
	\label{fig:squeezed}
\end{figure}

Following a similar procedure as in the previous sections, we can now compute the contribution to the density-weighted galaxy shape power spectrum of the two tensor non-Gaussianities discussed in the previous paragraph.
\begin{figure}[ tbp]
	\centering
	\includegraphics[width=0.495\textwidth]{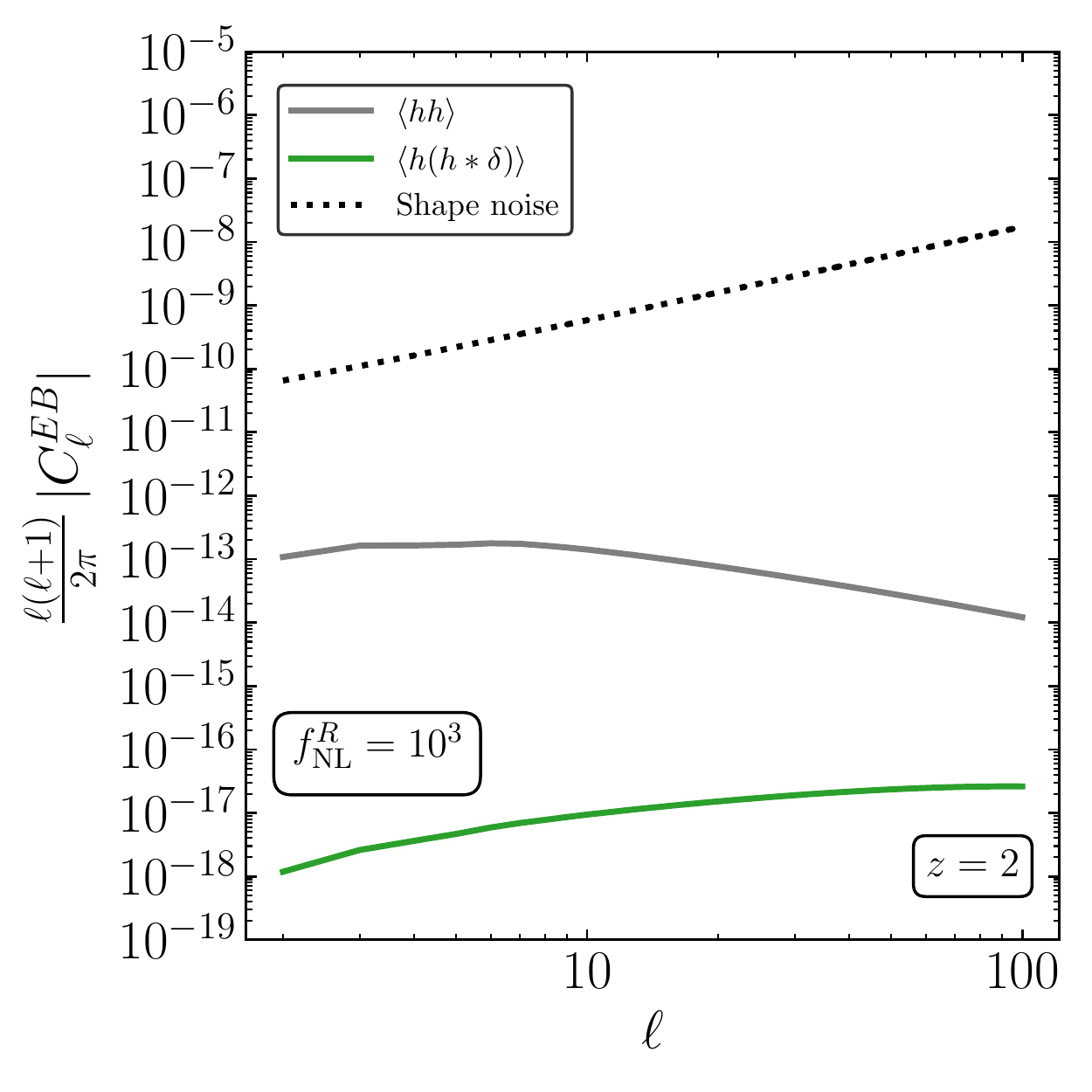}
	\includegraphics[width=0.495\textwidth]{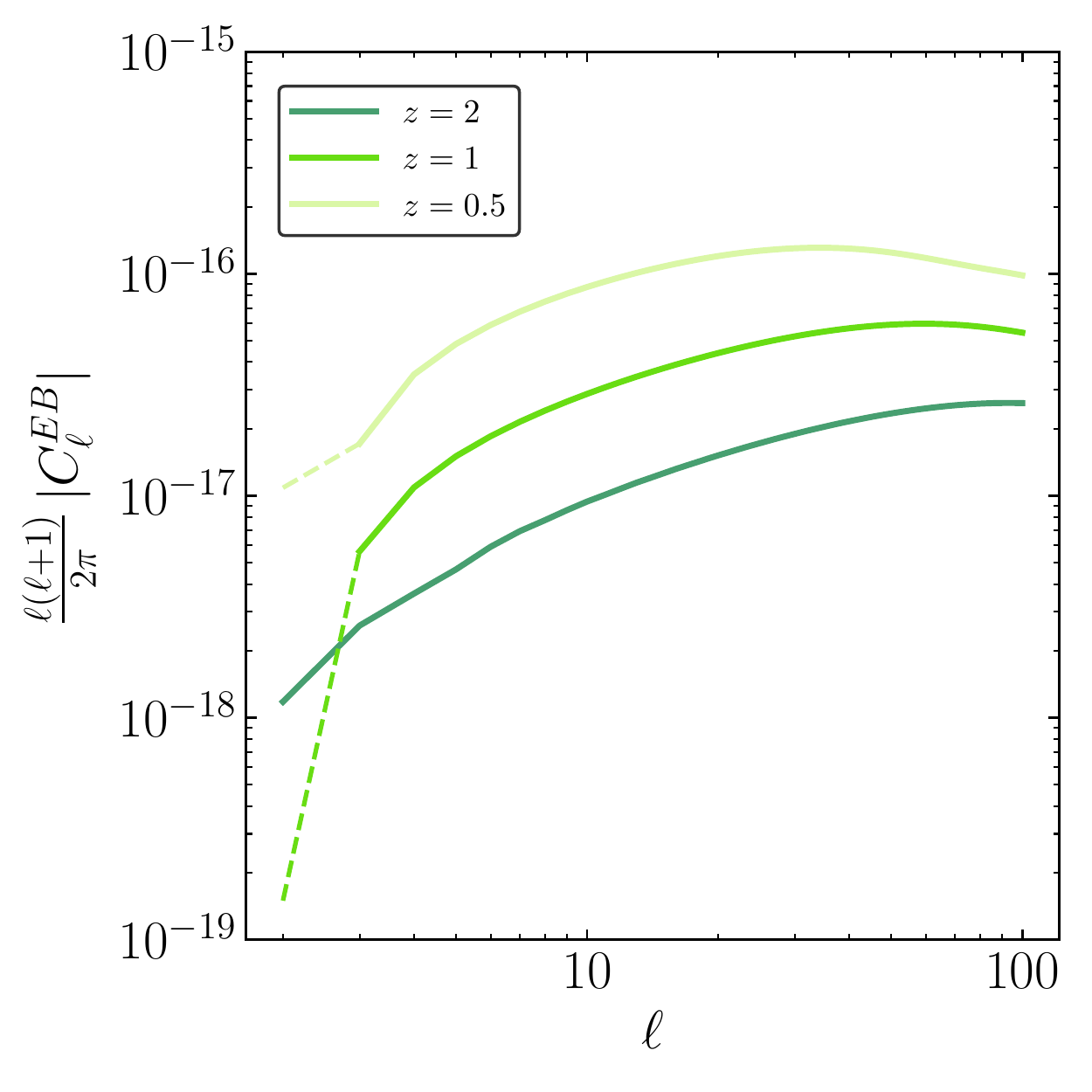}
	\caption{Left Panel: The contribution from the primordial parity-breaking bispectrum of Eq. \eqref{eq:hhzeta} (green solid) with the largest amplitude allowed by theoretical bounds, $f_{\rm NL}^R=10^3$, is compared to the power spectrum one (gray solid) and to the $1\sigma$ uncertainty of a LSST-like survey (black dotted line) at redshift $z=2$ . Right Panel: Contribution from the primordial parity-breaking bispectrum of Eq. \eqref{eq:hhzeta} for three different redshifts and $f_{\rm NL}^R=10^3$. }
	\label{fig:chiralbis}
\end{figure}
\paragraph{The $\langle h\, (h*\delta)\rangle$ term.} Using Eq. \eqref{eq:hhzeta} as a source to the fourth term in Eq. \eqref{eq:expintu}, we get the parity-breaking component
\begin{equation}
    C^{EB, (h h \delta)}_\ell = \sqrt{2}\,r\,f_{\rm NL}^{R}\, b_h^2\, b_\delta\,\int \frac{dk}{2\pi}\, k^2 \alpha(k)\,P_h(k)\, F^{E2}_\ell(x)\, F^{B2}_\ell(x)\, B_1(k) \, , 
\end{equation}
where here $r$ is the tensor to scalar ratio, not to be confused with the integration variable in the kernel
\begin{equation}\label{eq:bkernel}
    B_1(k) = \frac{k^3}{(2\pi)^2}\int_0^\infty dr\int_{-1}^{1}d\mu\, \alpha(r k)\, P_\zeta(k\sqrt{1+r^2-2 r \mu})\,\mathcal M(k\sqrt{1+r^2-2 r \mu}) \tilde B_1(r,\mu) \, ,
\end{equation}
being
\begin{equation}
    \tilde B_1(r,\mu) =  2(1+r)\,\mu\,(1-\mu^2)^2 \, .
\end{equation}
We have also reparameterized the amplitude of the primordial tensor bispectrum
\begin{equation}
    f_{\rm NL}^R \equiv \frac{25}{24576} \Pi
\end{equation}
as done in \cite{Bartolo:2017szm}. In the previous paragraph and in the Appendix \ref{squeezed_limit} we have discussed that the squeezed limit of Eq. \eqref{eq:hhzeta} should scale as $(k_L/k_S)^2$ after reabsorbing unphysical contributions. In our kernel, Eq. \eqref{eq:bkernel}, the squeezed limit corresponds to $r\rightarrow 1$ and $\mu \rightarrow -1$, which clearly gives a negligible contribution to the integrand. We have checked numerically that this is indeed the case. We therefore do not correct the squeezed limit as it does not affect our results.
Figure \ref{fig:chiralbis} illustrates the contribution of the parity-breaking primordial bispectrum of Eq. \eqref{eq:hhzeta} compared to the tensor power spectrum (left panel) and for different redshifts (right panel), where we again choose $b_\delta = b_h =1$. Following our discussion, detailed in Appendix \ref{bound:Pi}, we are using the maximum amplitude allowed for consistency of the theory, $f_{\rm NL}^R=10^3$. The results clearly show that it would be challenging to reach the necessary sensitivity to distinguish this contribution. 
\paragraph{ The $\langle h (K*\delta) \rangle$ term.} Using Eq. \eqref{eq:hzetazeta} as a source to the third term in Eq. \eqref{eq:expintu}, where we maximize the amplitude by assuming $\nu=3/2$, we get the B-mode component\footnote{An E-mode is also sourced by this non-Gaussianity, but we do not show it here as the scalar-induced E-mode galaxy shape power spectrum is quite large, cfr. Figure \ref{fig:scalarterms}.}
\begin{equation}\label{eq:cbbsque}
    C^{BB, (h K \delta)}_\ell = f_{\rm NL}^{h}\, b_h\,b_K\,b_\delta\, \int \frac{dk}{2\pi}\, k^2 \alpha(k)\,P_h(k)\, |F^{B2}_\ell(x)|^2\, \, B_2(k) \, ,
\end{equation}
where
\begin{equation}\label{eq:bkernel_squeezed}
    B_2(k) = \frac{k^3}{(2\pi)^2}\int_{0}^\infty dr\int_{-1}^{1}d\mu\, \, P_\zeta(r k)\,\mathcal M(r k)\,\mathcal M(k\sqrt{1+r^2-2 r \mu}) \tilde B_2(r,\mu) W_H\left(\frac{q}{k},R_{\rm squeezed}\right)\, ,
\end{equation}
being
\begin{equation}
    \tilde B_2(r,\mu) =  \frac14\, r^2\,(1-\mu^2) \, .
\end{equation}
and
\begin{equation}
    W_H(x,R) = 1-e^{-(x/R)^4}
\end{equation}
is introduced in order to ensure that we stay in the regime of validity of Eq. \eqref{eq:hzetazeta} by suppressing the integrand when the ratio of the short mode $q$ over the long mode $k$ is smaller than $R_{\rm squeezed} \gg 1$. In Figure \ref{fig:squeezed}, we show the contribution of the squeezed bispectrum to the total BB galaxy shape power spectrum at redshift $z=2$ for $R_{\rm squeezed}=100$ (left panel) and all the bias parameters are set to unity $b_K = b_\delta = b_h =1$ for similar reasons as argued above. We also verify the dependence on $R_{\rm squeezed} \in [10,100]$ of our results (right panel). Because we do not have the full shape, but only the squeezed limit of Eq. \eqref{eq:hzetazeta}, these results should be taken more as a rough order of magnitude estimation, rather than a precise prediction. Nevertheless, on the largest scales,  i.e. for $\ell\lesssim 20$ where the tensor-scalar-scalar bispectrum dominates the power spectrum, we expect the squeezed limit to hold. 
%However, the full tensor-scalar-scalar bispectrum shape would be needed to accurately compute the contribution. 
\section{Conclusions}\label{sec:conclusions}
In this paper, we have employed recently developed methods to get an insight into the physics of galaxy intrinsic alignments, focusing on the observational imprints from the early Universe. Our analysis pointed out three main aspects in this direction: first, we have calculated the contribution on E and B modes of the galaxy shape power spectrum sourced by the convolution of the shear field $K_{ij}$ with the matter density $\delta$ at large scales. This contribution is inevitable since the information on the shapes comes from the light emitted by an observable galaxy, thus weighted  by  the  galaxy  number  density  field. We found that the B-mode signal is generically stronger in amplitude, although different in shape, than the one sourced by gravitational waves of primordial origin (Fig. \ref{fig:tensorvsscalar}). This result suggests that searching for the imprint of primordial gravitational waves in B modes of the galaxy shape power spectrum is even more challenging than what stated in previous literature. 

Thus, realizing that such contamination would be absent if primordial gravitational waves broke parity, we have computed the EB galaxy shape power spectrum for a generic inflationary model predicting chirality in the primordial tensor power spectrum as in the form of Eq. \eqref{eq:chi}. We found that, although the expected signal is small (Fig. \ref{fig:chiral}), if detected it would represent a clear signature of parity breaking setups in the primordial Universe. In fact, we argue that the so-called shape noise signal, being invariant under parity transformations, should be vanishing for EB correlations.

Third, we point out that, similarly to what already studied for galaxy clustering statistics, the galaxy shape power spectrum is sensitive to primordial non-Gaussianities, in this case sourced by cross-correlations of tensors and scalar primordial perturbations. We computed the signature in the EB galaxy shape power spectrum from a parity-breaking tensor-tensor-scalar primordial bispectrum sourced by the gravitational Chern-Simons term during inflation. Even in our simplified setup, we find that observing such signatures with an idealized LSST-like survey would be challenging (Fig. \ref{fig:chiralbis}). Furthermore, we estimated the signature in the B modes galaxy shape power spectrum from squeezed tensor-scalar-scalar primordial bispectra that break consistency relations during inflation. We found that, on the largest scales, the expected signal overcomes the signature given by the power spectrum of primordial gravitational waves (Fig. \ref{fig:squeezed}).

Our study should be improved on a number of aspects: first of all, for a complete treatment, the EFT approach of \cite{Vlah:2019byq} would be a powerful tool for computing all possible contributions to the galaxy shape power spectrum. Even though we believe that our simplified scenario captures the essential qualitative features of the setup we studied, this should be checked. Second of all, we have estimated observability of our signatures with a rough, order of magnitude, evaluation of the expected stochastic noise. This should be complemented by more up-to-date models (see e.g. \cite{Chang:2013xja}) and by an estimation of the instrument systematics. In our specific case, it would be important to ascertain systematic uncertainties on E-B correlations. Finally, in our study of scalar-tensor primordial non-Gaussianities we have only considered imprints on the galaxy shape power spectrum, while the subsequent study of the galaxy shapes bispectra might provide a promising observational channel for more mixed scalar-tensor primordial non-Gaussianities and the full tensor bispectrum. We leave these studies for future research.

\paragraph{Acknowledgements}

We thank Nicola Bartolo, Daniel Baumann, Giovanni Cabass, Elisa Chisari, Sabino Matarrese, Guilherme Pimentel, Fabian Schmidt, Gianmassimo Tasinato and Zvonimir Vlah for helpful conversations and Daan Meerburg, Fabian Schmidt and Antonio Riotto for comments on a draft. G.\,O. would like to thank also the Institute of Physics of the University of Amsterdam for the warm hospitality during his visit in Amsterdam. M.\,B. acknowledges support from the Netherlands Organization for Scientific Research~(NWO), which is funded by the Dutch Ministry of Education, Culture and Science~(OCW), under VENI~grant~016.Veni.192.210. G.\,O. acknowledges partial financial support by ASI Grant 2016-24-H.0. and by the European programme Erasmus+ for traineeship. 

\appendix
\section{Details on the calculations of the galaxy shape power spectrum}\label{app:scalars}
In this Section we specify further details on computations we present in Sections  \ref{sec:scalar} and  \ref{sec:tensors}. While most of the formalism was developed in \cite{Schmidt:2012ne,Schmidt:2012nw}, we optimize computations to be fast and accurate both at large scales and small scales.
\subsection{More on the projection in the sky and useful relations}
As outlined in the main text, we can decompose $\tilde\gamma_{ij}$ into spin $\pm2$ functions as \cite{Schmidt:2012nw}
\be\label{eq:pm2}
{}_{\pm2}\tilde \g = \tilde \g_1 \pm i \tilde \g_2 \, .
\ee
This is analogous to the decomposition in Stokes parameters $Q\pm i U$ that is commonly used in the CMB linear polarization. Indeed, in analogy with the CMB case we will use multiple moments of E and B modes for the shear, which are invariant under a rotation of the frame and are eigenstates of parity.
%In particular, shear is quantitatively defined as the traceless part of the tensorial quantity $\A_{ij}$, which describes the transverse distortion of transverse standard rulers \cite{Pajer:2013ana}. A standard ruler is an observable defined in CFC coordinates. In our case the quantity $\A_{ij}$ describes the distortion of the cross-correlation of the ellipticities of the galaxies. 
Let us therefore define on the sphere the unit vectors of the circularly polarized basis, i.e.
$\v{m}_\pm \equiv (\v{e}_\theta\mp i\,\v{e}_\phi)/\sqrt{2}$. In this case the ${}_{\pm2}\tilde\g$ components of the shear are given by definition as
\ba
{}_{\pm2}\tilde\g \equiv\:& m_\mp^i m_\mp^j \tilde \gamma_{ij}  \, 
\ea
%and conversely the tensor $\gamma_{ij}$ can be decomposed on the sphere as
%\ba
%\tilde \g^{\rm IA}_{ij} =\:& {}_{+2}\g\, m_+^i m_+^j + {}_{-2}\g\, m_-^i m_-^j.
%\label{eq:sheardecomp}
%\ea
and represent spin-2 quantities in the sky. We can consider the contribution from a single plane wave to ${}_{\pm2}\tilde\g$ to find, in our LA model,
\begin{align}\label{eq:pmscalar}
    {}_{\pm2}\tilde\g (\hat \bn,\vk) =&\, b_K\, \frac{(m_\mp\cdot \vk)^2}{k^2}\, \delta(\vk,z_O)\,e^{i \vk\cdot\hat\bn \chi(z_O)}+ b_{\delta} \,b_{K}\, \int_{\vq}\frac{(m_\mp\cdot \vq)^2}{q^2}\, \delta(\vq,z_O)\,\delta(\vk-\vq,z_O)\,e^{i \vk\cdot\hat\bn \chi(z_O)}\nonumber\\
        &+ b_K\, b_2 \,\int_{\vq_1,\vq_2}\frac{(m_\mp\cdot \vq_1)^2}{q_1^2}\, \delta(\vq_1,z_O)\,\delta(\vq_2,z_O)\,\delta(\vk-\vq_1-\vq_2,z_O)\,e^{i \vk\cdot\hat\bn \chi(z_O)}\nonumber\\
    &+ b_K\, b_{K^2}\, \int_{\vq_1,\vq_2}\frac{(m_\mp\cdot (\vk-\vq_1-\vq_2))^2}{|\vk-\vq_1-\vq_2|^2}\left(\frac{(\vq_1\cdot\vq_2)^2}{q_1^2 q_2^2}-\frac 13\right)\nonumber\\
    &\phantom{+ b_K\, b_{K^2}\, \int_{\vq_1,\vq_2}}\times\delta(\vq_1,z_O)\,\delta(\vq_2,z_O)\,\delta(\vk-\vq_1-\vq_2,z_O)\,e^{i \vk\cdot\hat\bn \chi(z_O)} \, ,
\end{align}
where $\int_{\vq} \equiv \int d^3q / (2\pi)^3$ and notice that we are now explicitly indicating the dependence on the observed redshift $z_O$.
In order to compute observable correlations of such quantities, we can apply spin-lowering or -raising operators \cite{Schmidt:2012nw} to convert ${}_{\pm2}\tilde\g$ into a scalar in the sky
%. For this purpose, let us consider a single Fourier mode $\bk$, chosen to be along the $z$-axis, and apply the spin-lowering operator to \mb{refine}
\be \label{gamma_low}
\tilde\gamma(\vnhat,\vk) = \bar\Del^2\,{}_{+2}\tilde\g(\vnhat,\vk) = \left(-\frac{\partial}{\partial\mu} - \frac{m}{1-\mu^2}\right)^2 \left[ (1-\mu^2)\:{}_{+2}\tilde\g(\vnhat,\vk)  \right]\, ,
\ee
where $\mu$ is the cosine of the angle between $\vk$ and $\vnhat$ and $m = -2,-1,0,+1,+2$ depending on how the specific operator in the expansion of Eq. \eqref{eq:exptilde} transforms. As an example, for the operator $K_{ij}$ we get
\begin{align}
    \tilde \gamma(\vnhat,\vk) &\supset 
    %b_K \left(\frac{\partial}{\partial\mu}\right)^2\left[(1-\mu^2)\, m^i_-\, m^j_-\, \mathcal P_i^\ell\, \mathcal P_j^m\, K_{\ell m}(\k,z_O)\right]\nonumber\\
    -\frac 12\, b_K\, \left(\frac{\partial}{\partial\mu}\right)^2 \left[\delta(\vk,z_O)\,(1-\mu^2)^2\,e^{i x \mu}\right]\nonumber\\
    & = \frac 12\, b_K\, \delta(\vk,z_O)\, Q_0(x)\, e^{ix\mu}\, ,
\end{align}
where we defined $x= k \,\chi(z_O)$ and the derivative operator
\begin{equation}
    Q_0(x) = \left[4+12\partial_x^2+8x\partial_x+8x\partial_x^3+x^2+2x^2\partial_x^2+x^2\partial_x^4\right]\, .
\end{equation}
%Using the same procedure, the contribution to $\tilde \gamma$ from the other terms of Eq. \eqref{eq:pmscalar}, can be computed, as detailed in Appendix \ref{app:scalars}.
%Using the same procedure on the second term of the expansion Eq. \eqref{eq:exptilde}, we find the  expansion in terms of the scalar quantity
%\begin{equation}
%    \tilde \gamma(\vk,\vnhat) = \frac 12 b_K \delta(\vk,\vnhat) Q_0(x) e^{ix\mu} + ...,
%\end{equation}
%where we have defined derivative operators $Q_n(x)$ in appendix \ref{}. 
More in general, as shown in \cite{Schmidt:2012ne}, we can always turn derivatives with respect to $\mu$ into powers of $ix$ and powers of $\mu$ into derivatives with respect to $ix$. We then obtain the generic formula
\begin{equation}
    \bar\Del^2\,{}_{+2}\tilde\g(\vnhat,\vk) = \hat Q_i(x) (1-\mu^2)^{|r|/2}e^{ir\phi}e^{ix\mu}a(\vk)\, ,
\end{equation}
where now $a(\vk)$ is a scalar field and we define the derivative operators
\begin{align} 
\hat Q_{0}(x) &= 4 + x^2 + 8 x \, \partial_x + 12 \, \partial_x^2 + 2 x^2 \partial_x^2 + 8 x \partial_x^3 + x^2 \partial_x^4 \\
\hat Q_{-1}(x) &= i x^2 \partial_x^3 - x^2 \partial_x^2 + 8 i x \partial_x^2 +i  x^2 \partial_x - 4  x \partial_x +12 i \partial_x - x^2 + 4 i x\\
\hat Q_{1}(x) &= i x^2 \partial_x^3 + x^2 \partial_x^2 + 8 i x \partial_x^2 +i  x^2 \partial_x + 4  x \partial_x +12 i \partial_x + x^2 + 4 i x\\
\hat Q_{-2}(x) &= - x^2 \partial_x^2 - 2 i x^2 \partial_x -8 x \partial_x + x^2 - 8 i x -12 \\
\hat Q_{2}(x) &= - x^2 \partial_x^2 + 2 i x^2 \partial_x -8 x \partial_x + x^2 +8 i x -12 \, ,
\end{align}
which respect the following properties
\begin{align}
    \hat Q_n(x) &= (-1)^n \hat Q^*_{-n}(x)\\
    \hat Q^*_n(x) &=Q_n(-x)\, .
\end{align}
When writing down harmonic coefficients, we use the relation Eq. \eqref{harmonic_eq} to integrate over the angle analytically, obtaining as a result spherical bessel functions $j_\ell$ as a function of $x = k \chi$. 
The action of the derivatives operator $\hat Q(x)$ on these spherical Bessel functions gives
\begin{align} \label{Q:def}
Q_{0}(x) \, j_\ell(x) &= \frac{ (\ell-1) \ell (\ell+1) (\ell+2) j_{\ell}(x)}{x^2} \nonumber\\
Q_{-1}(x) \, \frac{j_\ell(x)}{x} &= - (-2 + \ell + \ell^2) \frac{j_{\ell}(x)}{x} + i \, (-2 + \ell + \ell^2) \left[(1+\ell) \frac{j_\ell(x)}{x^2} - \frac{j_{\ell+1}(x)}{x}\right] \nonumber\\
Q_{1}(x) \, \frac{j_\ell(x)}{x} &=  (-2 + \ell + \ell^2) \frac{j_{\ell}(x)}{x} + i \, (-2 + \ell + \ell^2) \left[(1+\ell) \frac{j_\ell(x)}{x^2} - \frac{j_{\ell+1}(x)}{x}\right] \nonumber\\
Q_{-2}(x) \, \frac{j_\ell(x)}{x^2} &= \frac{2 x j_{\ell+1}(x)-\left(\ell^2+3 \ell-2 x^2+2\right) j_\ell(x)}{x^2} + i \, \, \frac{2 x j_{\ell+1}(x)- 2(\ell+2) j_\ell(x)}{x} \nonumber\\
Q_{2}(x) \, \frac{j_\ell(x)}{x^2} &=  \frac{2 x j_{\ell+1}(x)-\left(\ell^2+3 \ell-2 x^2+2\right) j_\ell(x)}{x^2} - i \, \, \frac{2 x j_{\ell+1}(x)- 2(\ell+2) j_\ell(x)}{x} \,,
\end{align}
from which we define the appropriate transfer functions for E and B modes,
\begin{align}
    F^{E |r|}_\ell(x) &\equiv {\rm Re}\left[\hat Q_r (x)\right]\,\frac{j_\ell(x)}{x^{|r|}}\\
    F^{B |r|}_\ell(x) &\equiv {\rm Im}\left[\hat Q_r (x)\right]\,\frac{j_\ell(x)}{x^{|r|}}\, ,
\end{align}
where $r=0,\pm1,\pm2$.

\subsection{Scalar-induced contributions}
In Section  \ref{sec:scalar} we compute all the contributions up to order $\mathcal O(P_\delta^2)$ in the context of the LA model using the density-weighting galaxy shape field, $\tilde\gamma$. For completeness, we report here the expansion in terms of ${}_{\pm2}\tilde\gamma$ components
\begin{align}
    {}_{\pm2}\tilde\g (\hat \bn,\vk) =&\, b_K\, \frac{(m_\mp\cdot \vk)^2}{k^2}\, \delta(\vk,z_O)\,e^{i \vk\cdot\hat\bn \chi(z_O)}+ b_{\delta} \,b_{K}\, \int_{\vq}\frac{(m_\mp\cdot \vq)^2}{q^2}\, \delta(\vq,z_O)\,\delta(\vk-\vq,z_O)\,e^{i \vk\cdot\hat\bn \chi(z_O)}\nonumber\\
        &+ b_K\, b_2 \,\int_{\vq_1,\vq_2}\frac{(m_\mp\cdot \vq_1)^2}{q_1^2}\, \delta(\vq_1,z_O)\,\delta(\vq_2,z_O)\,\delta(\vk-\vq_1-\vq_2,z_O)\,e^{i \vk\cdot\hat\bn \chi(z_O)}\nonumber\\
    &+ b_K\, b_{K^2}\, \int_{\vq_1,\vq_2}\frac{(m_\mp\cdot (\vk-\vq_1-\vq_2))^2}{|\vk-\vq_1-\vq_2|^2}\left(\frac{(\vq_1\cdot\vq_2)^2}{q_1^2 q_2^2}-\frac 13\right)\nonumber\\
    &\phantom{+ b_K\, b_{K^2}\, \int_{\vq_1,\vq_2}}\times\delta(\vq_1,z_O)\,\delta(\vq_2,z_O)\,\delta(\vk-\vq_1-\vq_2,z_O)\,e^{i \vk\cdot\hat\bn \chi(z_O)}\, ,
\end{align}
from which, using Eq. \eqref{gamma_low}, \eqref{eq:spharm} and \eqref{harmonic_eq} we can get the harmonic coefficients,
\begin{align}
a_{\ell m}^{K}(\vk,\vnhat) = -\frac{1}{2}b_K \,\sqrt{4 \pi (2 \ell +1)} \, i^{\ell}  \sqrt{\frac{(\ell - 2)!}{(\ell +2)!}}\,  \, \delta(\bk, \eta)\, Q_0(x) \, j_{\ell}(x)\, ,
\end{align}  
%Using Eq. \eqref{eq:almeb} we can define the E and B harmonic coefficients
%\begin{align}
%    a_{\ell m}^{K, E}(\vk,\vnhat) &= -\frac{1}{2}c_K \,\sqrt{4 \pi (2 \ell +1)} \, i^{\ell}  \sqrt{\frac{(\ell - 2)!}{(\ell +2)!}}\,  \, \delta(\bk, \eta)\, Q_0(x) \, j_{\ell}(x),\\
%    a_{\ell m}^{K,B}(\vk,\vnhat) &= 0,
%\end{align}
%where we used the fact that $Q_0$ is a real function. The second term of Eq. \eqref{eq:pmscalar} gives
for the linear term and
\begin{align}
a_{\ell m}^{\delta K} (\vk,\vnhat) =&\,-\frac 12 b_K\,b_{\delta}\, \sqrt{4 \pi (2 \ell +1)} \, i^{\ell}  \sqrt{\frac{(\ell - 2)!}{(\ell +2)!}}\,  \int_{\bq}   \, \delta(\vq,\eta)\delta(\vk-\vq,\eta) \nonumber\\
&\left[\frac12 \, \delta_{m0}Q_{0}(x) \, j_\ell(x) \,  (3 \cos^2(\theta_q) -1) \right.\nonumber\\
&\,\,\left. +\,\sqrt{\frac{(\ell + 1)!}{(\ell - 1)!}} \,i \,  (\delta_{m+1}Q_{+1}(x)-\delta_{m-1}Q_{-1}(x)) \, \frac{j_\ell(x)}{x} \, \sin (\theta_q) \cos (\theta_q) \right. \nonumber\\
  & \,\,\left.  -\frac 14 \,\sqrt{\frac{(\ell + 2)!}{(\ell - 2)!}}  \, (\delta_{m+2}Q_{+2}(x)+\delta_{m-2}Q_{-2}(x)) \, \frac{j_\ell(x)}{x^2} \,  \sin ^2(\theta_q)\right]\, ,
\end{align} 
for the second term, where $\theta_q$ is the angle of $\vq$ with respect to the $z-$axis. 
Using a similar procedure, we can compute the remaining two terms to be
\begin{align}
a_{\ell m}^{\delta^2 K} (\vk,\vnhat) =&\, -\frac 12 b_K\,b_{\delta^2}\, \sqrt{4 \pi (2 \ell +1)} \, i^{\ell}  \sqrt{\frac{(\ell - 2)!}{(\ell +2)!}}\,  \int_{\bq_1,\bq_2}   \, \delta(\vq_1,\eta)\delta(\vq_2,\eta)\delta(\vk-\vq_1-\vq_2,\eta) \nonumber\\
&\left[ \frac14 \, Q_{0}(x)\delta_{m0} \, j_\ell(x) \,  (3 \cos(2\theta_{q_1}) +1) \right.\nonumber\\
&\,\,+\left. \,\sqrt{\frac{(\ell + 1)!}{(\ell - 1)!}} \,i \,  (\delta_{m+1}Q_{+1}(x)-\delta_{m-1}Q_{-1}(x)) \, \frac{j_\ell(x)}{x} \, \sin (\theta_{q_1}) \cos (\theta_{q_1}) \right. \nonumber\\
  & \,\,\left.  -\frac 14 \,\sqrt{\frac{(\ell + 2)!}{(\ell - 2)!}}  \, (\delta_{m+2}Q_{+2}(x)-\delta_{m-2}Q_{-2}(x)) \, \frac{j_\ell(x)}{x^2} \,\sin ^2(\theta_{q_1}) \right]  \, .
\end{align} 
and
\begin{align}
a_{\ell m}^{K^2 K} (\vk,\vnhat) =&\,  - \frac 14\, b_K\,b_{K^2}\,\sqrt{4 \pi (2 \ell +1)} \, i^{\ell}  \sqrt{\frac{(\ell - 2)!}{(\ell +2)!}}\, \int_{\bq_1,\bq_2}   \, \delta(\vq_1,\eta)\delta(\vq_2,\eta)\delta(\vk-\vq_1-\vq_2,\eta) \nonumber\\
&\left[ \, Q_{0}(x)\delta_{m0} \, j_\ell(x) \,  \, A(\vq_1,\vq_2,k) \right.\nonumber\\
&\,\,\left. \,\sqrt{\frac{(\ell + 1)!}{(\ell - 1)!}} \,i \,  (\delta_{m+1}Q_{+1}(x)-\delta_{m-1}Q_{-1}(x)) \, \frac{j_\ell(x)}{x} \, B(\vq_1,\vq_2,k) \right. \nonumber\\
  & \,\,\left.  -\,\sqrt{\frac{(\ell + 2)!}{(\ell - 2)!}}  \, (\delta_{m+2}Q_{+2}(x)-\delta_{m-2}Q_{-2}(x)) \, \frac{j_\ell(x)}{x^2} \,  \,C(\vq_1,\vq_2,k) \right]  \,,
\end{align} 
where 
\begin{align}
    A(\vq_1,\vq_2,k) =\, &  \frac{1}{8} \Big( 4 k^2+q_1^2+q_2^2 - 8 k q_1 \cos[\theta_{q_1}]+ 3 q_1^2 \cos[2 \theta_{q_1}] - 8 q_2 k \cos[\theta_{q_2}] + 3 q_2^2 \cos[2 \theta_{q_2}] \nonumber  \\
    & + 4 q_1 q_2 \cos[\theta_{q_1} - \theta_{q_2}]+ 4 q_1 q_2 \cos[\theta_{q_1} + \theta_{q_2}] - 4 q_1 q_2 \cos[\phi_{q_1} - \phi_{q_2}] \sin[\theta_{q_1}] \sin[\theta_{q_2}]  \Big) \nonumber\\
    & \times \Big(k^2 + q_1^2 + q_2^2 - 2 k q_2 \cos[\theta_{q_2}] - 
 2 q_1 \cos[\theta_{q_1}] (k - q_2 \cos[\theta_{q_2}]) \nonumber \\
& + 2 q_1 q_2 \cos[\phi_{q_1} - \phi_{q_2}] \sin[\theta_{q_1}] \sin[\theta_{q_2}]\Big)^{-1} \, ,
\end{align}
\begin{align}
    B(\vq_1,\vq_2,k) =\, & \frac{1}{2}\Big(k - q_1 \cos[\theta_{q_1}] - q_2 \cos[\theta_{q_2}]\Big) \Big(e^{i \phi_{q_1}} q_1 \sin[\theta_{q_1}] + 
   e^{i \phi_{q_2}} q_2 \sin[\theta_{q_2}]\Big)  \nonumber  \\
    & \times \Big(k^2 + q_1^2 + q_2^2 - 2 q_1 \cos[\theta_{q_1}] (k + q_2 \cos[\theta_{q_2}])  \nonumber  \\
   & - 
   2 q_2 (k \cos[\theta_{q_2}] + q_1 \cos[\phi_{q_1} - \phi_{q_2}] \sin[\theta_{q_1}] \sin[\theta_{q_2}])\Big)^{-1} 
\end{align}
and
\begin{align}
    C(\vq_1,\vq_2,k) =\, & \frac{1}{8} e^{-2i (\phi_{q_1} + \phi_{q_2})} \Big(e^{i \phi_{q_2}} q_1 \sin[\theta_{q_1}] + 
   e^{i \phi_{q_1}} q_2 \sin[\theta_{q_2}]\Big)^2  \nonumber  \\
    & \times \Big(k^2 + q_1^2 + q_2^2 - 2 q_1 \cos[\theta_{q_1}] (k + q_2 \cos[\theta_{q_2}])  \nonumber  \\
   & - 
   2 q_2 (k \cos[\theta_{q_2}] + q_1 \cos[\phi_{q_1} - \phi_{q_2}] \sin[\theta_{q_1}] \sin[\theta_{q_2}])\Big)^{-1} \, .
\end{align}
We can now decompose into E and B modes using Eq. \eqref{eq:almeb} and compute the $C_\ell$'s. We separate the integration over $d^3 q$ from that over $d^3k$ by defining kernel functions
\begin{align}
    S_n(k) & = \frac{k^3}{(2\pi)^2}\int_0^\infty dr \int_{-1}^{1}d\mu\, P(r k) P( k\sqrt{1+r^2-2 r \mu}, \eta) \tilde S_n(r,\mu)\\
    R(k) & = \frac{k^3}{(2\pi)^2}\int_0^\infty dr\,  P(r k) \tilde R(r)\, ,
\end{align}
where
\begin{align}
    \tilde S_{F_2}(r,\mu) &=  \frac{r(3\mu^2-1)(7\mu+3r-10r \mu^2)}{28(1+r^2-2 r \mu)}\, ,\\
    \tilde S_0(r,x) &=\frac{r^2}{4}\,(3\mu^2-1)\left(3\mu^2-2+\frac{3(r\mu-1)^2}{1+r^2-2r\mu}\right)\, ,\\
    \tilde S_1(r,\mu) &=  2\,r^2\,\frac{\mu\,(1-\mu^2)(2r\mu^2 -r -\mu)}{1+r^2-2 r \mu}\, ,\\
    \tilde S_2(r,\mu) &=  \frac{r^2}{8}\frac{(1-\mu^2)^2(1+2r^2 -2r\mu)}{1+r^2-2r\mu}\, .
\end{align}
When dealing instead with the integration over $k$, the transfer functions $F^{X|r|}_\ell$ introduce highly oscillatory terms, which can slow down integration. A schematic form of some of these integrals is 
\begin{equation}
    \mathcal I(\ell, \eta) = \int_0^\infty \, dk \, k^2 \,  w(k) \, |j_\ell(k \eta)|^2 \, ,
\end{equation}
for some weight $w(k)$. Using known mathematical approximations \cite{Tomaschitz2014HighindexAO}, we explain how to compute these integrals efficiently in Appendix \ref{app:approx}.

\subsection{Tensor-induced contributions}
In Section  \ref{sec:tensors} we compute the contributions from tensor perturbations to the density-weighted galaxy shape power spectrum. The expansion in terms of ${}_{\pm2}\tilde\gamma$ components
reads
\begin{align}
    {}_{\pm2}\tilde\g (\hat \bn,\vk) \supset &\, b_h\,\alpha(k, z_O)\,m_\mp^i m_\mp^j h_{ij}^{(0)} (\vk,z_O)\,e^{i \vk\cdot\hat\bn \chi(z_O)}\nonumber\\
    &+ b_h\,\alpha(k, z_O)\,b_{\delta} \int_{\vq}m_\mp^i m_\mp^j h_{ij}^{(0)}(\vq,z_O)\,\delta(\vk-\vq,z_O)\,e^{i \vk\cdot\hat\bn \chi(z_O)}\, ,
\end{align}
from which, using Eqs. \eqref{gamma_low}, \eqref{eq:spharm} and \eqref{harmonic_eq} we can get the harmonic coefficients,
\begin{align}
a_{\ell m}^{h}(\vk,\vnhat) = -\frac{1}{4}\,b_h\, \alpha(k)\,\sqrt{4 \pi (2 \ell +1)} \, i^{\ell}  \left[\delta_{m+2}h_{+2}(\bk)Q_{+2}(x)+\delta_{m-2}h_{-2}(\bk)Q_{-2}(x)\right] \, \frac{j_\ell(x)}{x^2} \, ,
\end{align}  
for the linear term and
\begin{align}
    a_{\ell m}^{\delta h}(\vk,\vnhat)&=\,\frac{1}{2\sqrt{2}}\, b_h b_\delta \,  \sqrt{4 \pi (2 \ell +1)} \, i^{\ell}  \sqrt{\frac{(\ell - 2)!}{(\ell +2)!}}\,  \int_{\bq}  \, \alpha(q) \, \delta(\vk-\vq,\eta) \nonumber\\
&\left[ \frac 32\, \delta_{m0}Q_{0}(x) \, j_\ell(x) \,\left[h_{+2}(\vq)+h_{-2}(\vq)\right] \, \sin^2(\theta_q) \right.\nonumber\\
&- \left. \,\sqrt{\frac{(\ell + 1)!}{(\ell - 1)!}} \,i \,  \delta_{m+1}Q_{+1}(x) \, \frac{j_\ell(x)}{x} \, \sin (\theta_q)\, \sum_{ p=+1,-1} (\cos (\theta_q) + p )h_{2p}(\vq) \right. \nonumber\\
&-\left. \,\sqrt{\frac{(\ell + 1)!}{(\ell - 1)!}} \,i \,  \delta_{m-1}Q_{-1}(x) \, \frac{j_\ell(x)}{x} \, \sin (\theta_q) \sum_{ p=+1,-1} (\cos (\theta_q) -p )h_{2p}(\vq) \right. \nonumber\\
&\left.  \, +\frac 14\, \sqrt{\frac{(\ell + 2)!}{(\ell - 2)!}}\delta_{m+2}Q_{+2}(x) \, \frac{j_\ell(x)}{x^2} \, \sum_{ p=+1,-1}(\cos\theta_q + p)^2\, h_{2p}(\vq) \right. \nonumber\\
&\left. \, +\frac 14\,\sqrt{\frac{(\ell + 2)!}{(\ell - 2)!}}\delta_{m-2}Q_{-2}(x) \, \frac{j_\ell(x)}{x^2} \, \sum_{p=+1,-1} (\cos\theta_q - p)^2\, h_{2p}(\vq)\right]\, ,
\end{align} 
for the second term.
We can now decompose into E and B modes using Eq. \eqref{eq:almeb} and compute the $C_\ell$'s. As for the scalar-induced terms, we can make the numerical computation faster, by separating the integration over $d^3 q$ from that over $d^3k$ by defining kernel functions
\begin{align}
    T_n(k) & = \frac{k^3}{(2\pi)^2}\int_0^\infty dr \int_{-1}^{1}d\mu\, \alpha^2(r k)\, P_h(r k) P_\delta( k\sqrt{1+r^2-2 r \mu}) \tilde T_n(r,\mu)\, ,
\end{align}
where
\begin{align}
    \tilde T_0(r,\mu) &= \frac{3}{16}\, r^2\,(1-\mu^2)\, ,\\
    \tilde T_1(r,\mu) &= \frac 14\, r^2\, (1-\mu^4)\, ,\\
    \tilde T_2(r,\mu) &= \frac{1}{16} \, r^2\, (1+6\mu^2+\mu^4)\, .
\end{align}
We use similar techniques as the previous section to deal with the numerical integration.

\section{Overview of parity breaking models}\label{app:chiralmodels}

A first way of writing down models that break parity during inflation typically relies on effective field theories of inflation (see e.g. \cite{Weinberg:2008hq,Creminelli:2014wna, Baumann:2015xxa}), where new parity violating operators in the action of the theory are considered without adding new field content. In particular, it is possible to show that the only two independent parity breaking operators that we can build with the lowest number of derivatives are \cite{Creminelli:2014wna}
\begin{equation} \label{P_break}
\epsilon^{ijk} \, \partial_i \dot h_{jl} \, \dot h_{lk} \,\, , \qquad \quad   \epsilon^{ijk} \,  \partial_i \partial_m h_{jl} \, \partial_m h_{lk} \, .
\end{equation}
During inflation we can couple these operators with generic functions of the inflaton field $f_i(\phi)$ and get the effective parity breaking action
\begin{equation} \label{S_par_break}
S_{\cancel{\mathcal P}} = \int d^4 x \, a^3 \, \left[\frac{f_1(\phi_0)}{\Lambda} \,  \frac{1}{a}  \, \epsilon^{ijk} \, \partial_i \dot h_{jl} \, \dot h_{lk} - \frac{f_2(\phi_0)}{\Lambda} \frac{1}{a^3} \, \epsilon^{ijk} \,  \partial_i \partial_m h_{jl} \, \partial_m h_{lk}\right] \, ,  
\end{equation}
where $f_{1/2}(\phi_0)$ are generic dimensionless coupling functions with the inflaton field and the scale factors are given by the fact that we are in a (quasi)-de Sitter background space. Moreover, we must include an UV cut-off scale $\Lambda$ in the denominator of each term, which signals the scale at which the effective field theory is broken. In the context of inflation, this scale is supposed to be bigger than the characteristic energy scale of inflation, i.e. $\Lambda > H$. In fact, we are interested to study only those scales that go outside the Hubble horizon during inflation.

It is possible to show that the additional terms in the quadratic action \eqref{S_par_break} induce an opposite correction to the power spectrum of primordial gravitational waves with opposite helicities. In the super-horizon limit, this correction does not depend on $f_1$, but only on $f_2$, and reads \cite{Creminelli:2014wna}
\begin{equation}
\Delta P_{R/L}(k_*) = \pm \frac{\pi}{4} f_2(\phi_0^*) \frac{H_*}{\Lambda} \, P_h(k_*) \, ,
\end{equation}
where the star means that the parameters are evaluated at the time of horizon crossing of the wave-number $k_*$, and $P_h(k_*)$ is the total tensor power spectrum as given in Eq. \eqref{P_tensors}.

Thus, at linear level and at the lowest order in the derivatives all the effective field theory models of inflation introducing parity breaking signatures are expected to provide the following value of $\chi$ at a given scale $k_*$:
\begin{equation} \label{chi_general}
\chi(k_*) =  \frac{\pi}{2} f_2(\phi_0^*) \frac{H_*}{\Lambda}\, ,
\end{equation}
which in principle may be degenerate among the different models. However, due to the condition $\Lambda > H$ during inflation, the final amount of chirality produced within these effective models is expected to be far from the unity, i.e. $|\chi| \ll 1$.

In literature, in the context of effective field theory of inflation, the first parity breaking models proposed have been slow-roll models in presence of the 4-dimensional Chern-Simons modified gravity term. This theory consists in a parity-breaking modification of Einstein gravity in which the so-called Chern-Simons gravitational term coupled to the inflaton field is added in the action of the slow-roll inflationary models. This term can be written in terms of the Riemann tensor as 
\begin{equation}\label{Chern-Simons}
S_{R \tilde R} = \int d^4 x \left[ f(\phi)  \, \epsilon^{\mu\nu\rho\sigma} {R_{\mu \nu}}^{\kappa \lambda}R_{\rho \sigma\kappa\lambda} \right]\, ,
\end{equation}
where $f(\phi)$ is a generic function of the scalar inflaton field $\phi$ only, and $\epsilon^{\mu\nu\rho\sigma}$ is the total antisymmetric Levi-Civita pseudo-tensor. Effects of this kind of gravity on  primordial gravitational waves have been studied for the first time in \cite{Lue:1998mq}, while more recent works include \cite{Jackiw:2003pm,Alexander:2004wk, Satoh:2008ck,Alexander:2009tp, Satoh:2010ep,Myung:2014jha, Alexander:2004us, Alexander:2007qe, Noorbala:2012fh,Bartolo:2017szm, Kawai:2017kqt,Mylova:2019jrj}. Notice that, despite the fact that 4-dimensional Chern-Simons term is a 4-derivatives term, it is a total derivative term, thus 1 derivative can be integrated by parts and act on the coupling function $f(\phi)$. This fact allows to reconcile with the form \eqref{S_par_break} at quadratic level in tensor modes.

Another parity breaking model which shows the same pattern as \eqref{S_par_break} is slow-roll inflation in presence of the 3-dimensional Chern-Simons term, which is included in the context of quantum Horava-Lifschitz gravity \cite{Takahashi:2009wc, Wang:2012fi}.

Recently, in \cite{Crisostomi:2017ugk}, other scalar-tensor parity breaking operators which lead to the same pattern as \eqref{S_par_break} have been considered (see e.g. \cite{Qiao:2019hkz}). These operators are built by contracting the Riemann tensor with covariant derivatives of the inflaton field.

A second approach to obtain parity breaking signatures during inflation consists in adding in the theory gauge bosons coupled to a pseudo-scalar axion-like field through a Chern-Simons like operator. In particular, the toy model action of these theories reads
\begin{equation} \label{Chromo}
S_{Chromo} = \int d^4 x \left[ -\frac{1}{4} F_{\mu \nu}^a F^{\mu \nu}_a + \frac{\lambda \phi}{4 f} F_{\mu \nu}^a \tilde F^{\mu \nu}_a\right]  \, , 
\end{equation}
where $\lambda$ and $f$ are respectively dimensionless and dimensionful constants, $\phi$ denotes the inflaton field which is a pseudo-scalar (axion), $F^{\mu \nu}_a$ is the field strength of a certain vector gauge field $A_{\mu}^a$ with the index $a$ transforming under the Lie group $\mathcal G$  algebra, i.e. $F^{\mu \nu}_a = \partial_\mu A^a_{\nu} - \partial_\nu A^a_{\mu} - g f^{abc} A^b_{\nu} A^c_{\nu}$ ($f^{abc}$ denote the structure constants of the algebra), and $ \tilde F^{\mu \nu}_a$ is its dual. 

The case of $\mathcal G = U(1)$ is known as pseudo-scalar inflation and it was the first parity breaking scenario based on action \eqref{Chromo} (see e.g. \cite{Sorbo:2011rz,Anber:2012du,Barnaby:2010vf,Barnaby:2012xt,Peloso:2016gqs}). However, observational data regarding the statistics of scalar perturbations put severe restrictions on the model \cite{Namba:2015gja}. The production of chirality is compatible with data only for certain specific wavenumbers of GW.

Thus, alternative scenarios have been considered, as for instance the chromo-natural inflation (CNI) scenario where the Lie group of the gauge field is $\mathcal G = SU(2)$ (see e.g. \cite{Maleknejad:2011jw, Dimastrogiovanni:2012ew, Adshead:2013nka, Adshead:2013qp, Maleknejad:2016qjz,Domcke:2018rvv,Watanabe:2020ctz}) and scenarios where the Chern-Simons term in \eqref{Chromo} is coupled to an external scalar field $\chi \neq \phi$ (see e.g. \cite{Dimastrogiovanni:2016fuu,McDonough:2018xzh,Maleknejad:2018nxz,Papageorgiou:2019ecb}), allowing to relax some issues of the pseudo-scalar scenario.

All the models characterized by this second approach rely on a mechanism of amplification of only one of the two chiral modes of gravitational waves due to the appearing of a source term  experiencing a tachionic growth during inflation. Differently by models based on action \eqref{S_par_break}, in this class of models we can get $P_R \gg P_L$ (or vice-verse), thus $|\chi| \simeq 1$.

\section{Bound on the amplitude of tensor-tensor-scalar chiral bispectrum} \label{bound:Pi}

A theoretical constraint on the value of $\Pi$ (Eq. \eqref{Pi}) occurs when we look to the radiative stability of the Chern-Simons modified gravity theory. The bispectrum \eqref{eq:hhzeta} comes from the following tensor-tensor-scalar interaction vertex (in Fourier space) \cite{Bartolo:2017szm}
\begin{equation}\label{eq:L_stt}
L_{int}^{ h h \delta \phi} = - \lambda_s  \, \int d^3 K  \frac{\delta^3(\vec k+ \vec p+ \vec q)}{(2 \pi)^6}  \left[\, a \left(\dot \phi \frac{\partial^2 f(\phi)}{\partial^2 \phi} \right) p \left( \vec{p} \cdot \vec{q} \right) \mbox{ } h^s_{ij} (\vec p) h^{s,\, ij} (\vec q) \delta \phi(\vec k) \right] \, , 
\end{equation}
where $ \int d^3 K= \int d^3k \mbox{ }d^3p \mbox{ }d^3q$ and  ${h}^s_{ij}(\vec p)={h}_s(\vec p) \, {\epsilon}^s_{ij}(\vec p)$, where ${\epsilon}^s_{ij}(\vec p)$ is the polarization tensor and ${h}_s(\vec p)$ is the graviton mode function. The Latin indices contractions are made with $\delta^{ij}$ and the primes $'$ indicate derivatives with respect to conformal time. The coefficient $\lambda_s$ takes $+1$($-1$) for R (L) polarization modes and the sum over the polarization index $s=R,L$ is understood for simplicity of notation.

Following the same reasoning of \cite{Baumann:2011su}, we switch to the following canonically normalized gravitons in de Sitter space
\begin{equation}
 h^s_{c}(\vec k)  =  \left(\frac{M_{Pl}^2}{2}\right)^{1/2} h^s(\vec k)  \, ,
\end{equation}
and we rewrite in terms of canonically normalized fields the interaction Lagrangian \eqref{eq:L_stt}. We obtain
\begin{align}
\mathcal L_{int}^{h h \delta \phi} =& -\lambda_s \, \int d^3 K \, \frac{\delta^3(\vec k+ \vec p+ \vec q)}{(2 \pi)^6} \,  a \frac{2}{M^2_{Pl}} \, \left(\dot \phi \frac{\partial^2 f(\phi)}{\partial^2 \phi} \right) p \left( \vec{p} \cdot \vec{q} \right)\,  h^s_c(\vec p) h^s_c(\vec q) \delta \phi_c(\vec k) \, \epsilon^s_{ij} (\vec p) \epsilon_s^{ij} (\vec q) \nonumber\\
=& -\lambda_s \, \int d^3 K \, \frac{\delta^3(\vec k+ \vec p+ \vec q)}{(2 \pi)^6} \,  a\frac{1}{\Lambda_S^2} \, p \left( \vec{p} \cdot \vec{q} \right)\,  h^s_c(\vec p) h^s_c(\vec q) \delta \phi_c(\vec k) \, \epsilon^s_{ij} (\vec p) \epsilon_s^{ij} (\vec q) \, ,
\end{align}
where we defined
\begin{equation}
\Lambda_S^2 = \frac{M_{Pl}^2}{2} \, \left(\dot \phi \frac{\partial^2 f(\phi)}{\partial^2 \phi} \right)^{-1} \, .
\end{equation}
To avoid a strong coupling regime on super-horizon scales (which would spoil the perturbativity of the theory), we must impose 
\begin{equation}
H^2< \Lambda_S^2  \, ,
\end{equation}
which gives the following constraint on the strength of the second order derivative of the coupling function $f(\phi)$:
\begin{equation}
H^2 \frac{\partial^2 f(\phi)}{\partial^2 \phi}  < \frac{M_{Pl}}{H} \frac{1}{2 \sqrt{2 \epsilon}} \simeq \frac{2}{\sqrt 2} \left(\frac{0.1}{r} \right) \times  10^5 \, .
\end{equation}
Thus, recalling the definition of $\Pi$, Eq. \eqref{Pi}, we get the theoretical constraint 
\begin{equation} \label{bound_Pi}
\Pi  \lesssim  \left(\frac{0.1}{r} \right) \times  10^6  \, .
\end{equation}
From Eq. \eqref{bound_Pi}, it would seem that, by decreasing $r$, one would get less stringent bounds on $\Pi$. In reality, one should keep in mind that the bispectrum of Eq. \eqref{eq:hhzeta} is proportional to $r^2$ coming from the 2 tensor power spectra, attenuating the power of the bispectrum in the $r \rightarrow 0$ limit independently by the value of $\Pi$ (which is determined by the strength of the second order derivative of the coupling function $f(\phi)$).

\section{Squeezed limit of the tensor-tensor-scalar chiral bispectrum}
\label{squeezed_limit}

In this Appendix we comment on the the physical squeezed limit of our bispectrum \eqref{eq:hhzeta}, which corresponds to the limit in which the momentum of the scalar perturbation $\zeta$ is much smaller than the momenta of the two gravitons. In fact, it is well known from the literature (see \cite{Tanaka:2011aj,Dai:2013kra,Creminelli:2013mca,Creminelli:2013cga,Pajer:2013ana,Hinterbichler:2013dpa,Mirbabayi:2014zpa,Dai:2015rda,Dai:2015jaa,Bordin:2016ruc,Cabass:2016cgp,Cabass:2018roz}) that primordial bispectra usually contain unphysical contributions in the squeezed limit. Physically, in this limit we are taking the cross-correlation in the space between two fields evaluated at close points $\mathbf x_1$  and $\mathbf x_2$, and a third field evaluated at a point $\mathbf x_3$ that is far away to the infinite. It is possible to show that the physical signal of this cross-correlation is the one computed in the so-called Conformal Fermi Coordinate (CFC) frame centered in the point $\mathbf x_0$  which stays in the middle of $\mathbf x_1$ and $\mathbf x_2$ (see e.g. \cite{Pajer:2013ana}). This local frame is constructed by imposing that  the metric becomes unperturbed FRW along the time-like geodesic  passing through $\mathbf x_0$ (the so-called \textit{central geodesic}), with corrections that go as the spatial distance from the central geodesic squared and involve second order derivatives  of metric perturbations, as we would expect by the virtue of the equivalence principle.

Considering the bispectrum $B^{R/L}_{hh \zeta}(k_1, k_2, k_3)$ (Eq. \eqref{eq:hhzeta}), it is possible to show that in the squeezed limit where $k_3=k_L \ll k_1 \simeq k_2 \simeq k_S$ the leading order effects of the long-wavelength perturbation $k_L$ on the short modes can be removed by transforming to the CFC local frame, leaving only contributions starting from the order $(k_L/k_S)^2$. In particular, in co-moving gauge our physical squeezed bispectum up to order $(k_L/k_S)^2$ reads (see e.g. \cite{Pajer:2013ana})
\begin{align} \label{squeezed_master2}
B_{h h \zeta}^{R/L}(k_S, k_S, k_L)_{\rm squeezed, ph} = &\left[ \frac{d \log (k_S^3 P_h^{R/L}(k_S))}{d \log k_S} P_\zeta(k_L) P_h^{R/L}(k_S) \right. \nonumber \\
& \left. + B_{h h \zeta}^{R/L}(k_1, k_2, k_3)_{\rm squeezed}^{\vec k_1 = \vec k_S - \frac{1}{2} \vec k_L, \,\,\, \vec k_2 = - \vec k_S - \frac{1}{2} \vec k_L , \,\,\, \vec k_3 = \vec k_L} \right] \nonumber \\
& + \mathcal O\left(\frac{k_L}{k_S}\right)^2 \, ,
\end{align}
where $B_{h h \zeta}^{R/L}(k_1, k_2, k_3)_{\rm squeezed}$ denotes the bispectrum in global coordinates, and all the power spectra are computed at the time of the horizon crossing of the short momenta $k_S$. In particular, $P_h^{R/L}(k)$ is the R/L-handed tensor power spectrum of the form Eq. \eqref{eq:prl} and $P_\zeta(k)$ is the scalar power spectrum
\begin{equation}\label{pow_zeta}
P_\zeta(k) = \frac{2 \pi^2}{k^3} \mathcal A_s\, ,
\end{equation}  
where we neglected for simplicity the scalar tilt, taking $n_s - 1 \approx 0$. 

On super-horizon scales, we can rewrite the first term on the r.h.s. of Eq. \eqref{squeezed_master2} in terms of the  derivative with respect to the cosmological time as
\begin{equation}\label{term_subtr}
\frac{d \log (k_S^3 P_h^{R/L}(k_S))}{d \log k_S} P_\zeta(k_L) P_h^{R/L}(k_S) =  P_\zeta(k_L) \left(3 P_h^{R/L}(k(t)) + \frac{1}{H} \frac{d}{dt} P_h^{R/L}(k(t))\right)\Big|_{t=t_S}\, ,
\end{equation} 
where we used the fact that
\begin{equation}
d \log (k_S^3 P_h^{R/L}(k_S)) = 3 (d \log k_S) + \frac{d P_h^{R/L}(k(t_S))}{P_h^{R/L}(k(t_S))} \, ,
\end{equation}
and that each short mode $k_S$ can be related to the time $t_S$ of horizon crossing by the relation
\begin{equation}
k_S = k(t_S) = a(t_S) H(t_S) \, .
\end{equation}
In fact, since during quasi-de Sitter inflation $a \sim e^{H t}$, then we have (apart for slow-roll corrections)
\begin{equation}
d \log k_S = H dt_S \, .
\end{equation}
Computing the time derivative term in Eq. \eqref{term_subtr}, and taking only the contribution coming from the Chern-Simons modified gravity (the one which depends on the coupling function $f(\phi)$), we find
\begin{equation}
P_\zeta(k_L) \frac{1}{H} \frac{d}{dt_S} P_h^{(R/L)}(k_S) = \mp \frac{\pi}{8} r \, H^2 f''(\phi) \, P_\zeta(k_L) P_h(k_S) = \mp \frac{25}{768} r \, \Pi  \, P_\zeta(k_L) P_h(k_S) \, ,
\end{equation}
where in the last step we have used the definition of $\Pi$, Eq. \eqref{Pi}. 

On the other hand, the mathematical squeezed limit value of bispectrum \eqref{eq:hhzeta} reads
\begin{equation} \label{math:squeezed}
B^{R/L}_{h h \zeta}(k_S, k_S, k_L)|_{\rm squeezed} = \pm \frac{25}{768} r \, \Pi \, P_\zeta(k_L) P_h(k_S) \left[1- \left(\frac{k_L}{k_S}\right)^2 \right] + \mathcal O\left(\frac{k_L}{k_S}\right)^3 \, .
\end{equation}
Thus, as we would expect the first term on the r.h.s. of Eq. \eqref{squeezed_master2} exactly cancels out the leading order value of \eqref{math:squeezed}, leaving
\begin{equation} \label{phys_squez}
B^{R/L}_{h h \zeta}(k_S, k_S, k_L)|_{\rm squeezed, ph} = \mp \frac{25}{768} r \, \Pi \, \left(\frac{k_L}{k_S}\right)^2 \, P_\zeta(k_L) P_h(k_S) + \mathcal O\left(\frac{k_L}{k_S}\right)^3\, ,
\end{equation}
which can be rewritten using Eqs. \eqref{P_tensors} and \eqref{pow_zeta} as
\begin{equation} \label{phys_squez2}
B^{R/L}_{h h \zeta}(k_S, k_S, k_L)|_{\rm squeezed, ph} = \mp \frac{25 \pi^4}{192} \mathcal A_s^2 \, (r^2\Pi) \, \left(\frac{k_L}{k_S}\right)^2 \, \left(\frac{1}{k_L^3 k_S^3}\right) + \mathcal O\left(\frac{k_L}{k_S}\right)^3\,  .
\end{equation}
However, Eq. \eqref{phys_squez2} does not give the exact physical bispectrum up to and including the order $(k_L/k_S)^2$. In fact, in Eq. \eqref{squeezed_master2} we neglected those terms of order $(k_L/k_S)^2$ coming from the transformation of the bispectrum from global to local coordinates. In general, these additional terms provide a renormalization of the $\mathcal O({k_L}/{k_S})^2$ term in Eq. \eqref{phys_squez2} (see \cite{Cabass:2016cgp} for a computation of these terms in the case of the scalar bispectrum in standard gravity). In our case, we are not very sensitive to the exact expression of the physical bispectrum in the squeezed limit, as we verified that the contribution to the integral of Eq. \eqref{eq:cbbsque} coming from the squeezed configurations is negligible due to the $\propto (k_L/k_S)^2$ behaviour.

\section{High-$\ell$ approximation of integrals involving Spherical Bessel functions}\label{app:approx}

\begin{figure}[tbp]
	\centering
	\includegraphics[width=0.65\textwidth]{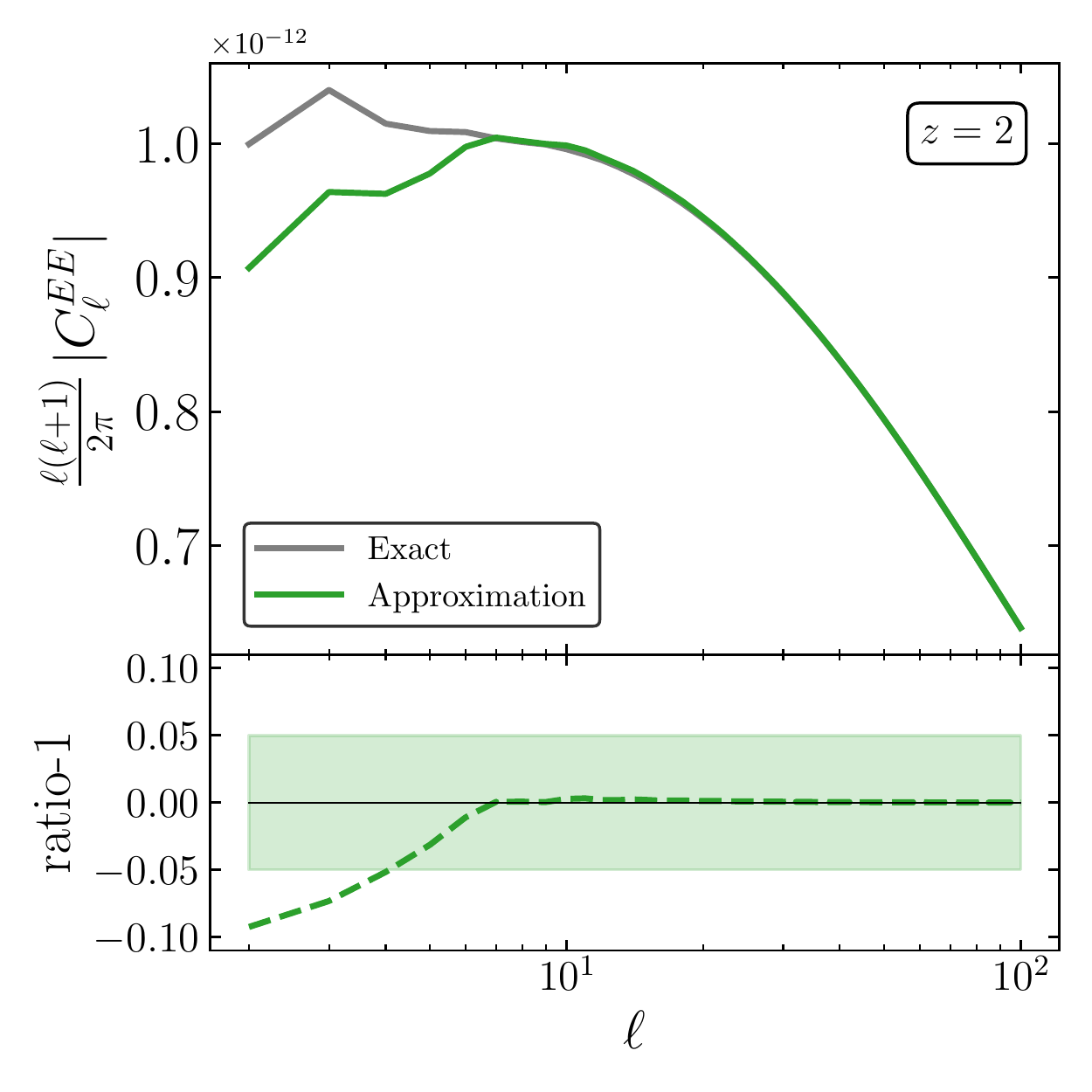}
	\caption{Upper panel: Contribution from tensor perturbations to the EE galaxy shape power spectrum at redshift $z=2$ for the exact (gray solid) and approximate (green solid) calculations. Bottom panel: Relative difference between exact and approximated results.}
	\label{fig:approx}
\end{figure}

In this work we deal with integrals of products of spherical Bessel functions $j_{\ell}(x)$ and $j_{\ell+1}(x)$ coming from the $F_{\ell}^{X|r|}$ transfer functions (see e.g. Eqs. \eqref{eq:hdhdE} and \eqref{eq:hdhdB}). However, all these integrals can be expressed in terms of products of a spherical Bessel function $j_{\ell}(x)$ of a given order $\ell$ and its derivative using the following recursive relation 
\begin{equation}
j_{\ell}'(x) = - j_{\ell+1}(x) + \left(\frac{\ell}{x}\right) j_{\ell}(x)\, .
\end{equation}
Thus, in this Appendix we provide formulae to approximate integrals involving products of spherical Bessel functions and their derivatives evaluated at high multipoles, adapting results from \cite{Tomaschitz2014HighindexAO}. These approximate formulae can be considered as a generalization of the flat-sky approximation to the transfer functions $F_{\ell}^{X|r|}$. This is motivated by the fact that spherical Bessel functions have an highly oscillatory behaviour which makes their numerical integration very inefficient. As an example, we have computed the contribution from tensor perturbations on the EE galaxy shape power spectrum, Eq. \eqref{eq:bbpower}, using the exact and approximated formulae, with results shown in Figure \ref{fig:approx}. The running time on a laptop with $2$ CPU cores is respectively $\sim 1$ minute and $\sim 5$ seconds and the approximation proves to be accurate to below $5\%$ for $l\gtrsim 5$.
We need to consider the following integrals:
\begin{align}
I_1(\ell, \eta) =& \int_0^\infty \, dk \, k^2 \,  w(k) \, |j_\ell(k \eta)|^2  \nonumber \\
I_2(\ell, \eta) =& \int_0^\infty \, dk \, k^2 \,  w(k) \, |j'_\ell(k \eta)|^2   \nonumber \\
I_3(\ell, \eta) =& \int_0^\infty \, dk \, k^2 \,  w(k) \, j_\ell(k \eta) \, j'_\ell(k \eta) \, , \label{integrals_Bessel}    
\end{align}
where $w(k)$ is a generic kernel function.
In the high-$\ell$ limit, integrals \eqref{integrals_Bessel} can be approximated by
\begin{align}
I_1(\ell, \eta)|_{\ell \rightarrow \infty} \simeq&  \frac{(\ell+1/2)}{4 \eta^3} \int_0^\infty \, \frac{dy}{\sqrt y} \,  w\left(\frac{(\ell+1/2)}{\eta} \sqrt{1+y}\right)  \nonumber\\
I_2(\ell, \eta)|_{\ell \rightarrow \infty} \simeq&  \frac{(\ell+1/2)}{4 \eta^3} \int_0^\infty \, dy \, \frac{\sqrt y}{1+y} \,  w\left(\frac{(\ell+1/2)}{\eta} \sqrt{1+y}\right) \nonumber \\
I_3(\ell, \eta)|_{\ell \rightarrow \infty} \simeq& - \frac{1}{4 \eta^3} \int_0^\infty \,  \frac{dy}{\sqrt y \sqrt{1+y}} \,  w\left(\frac{(\ell+1/2)}{\eta} \sqrt{1+y}\right) \nonumber \\
&- \frac{(\ell+1/2)}{8 \eta^4} \int_0^\infty \, \frac{dy }{\sqrt y} \,  w'\left(\frac{(\ell+1/2)}{\eta} \sqrt{1+y}\right) \, . \label{integrals:approx}
\end{align}
In the following, we will provide the derivation of these formulae.

\subsection{High-$\ell$ approximation of $\int_0^\infty \, dk \, k^2 \,  w(k) \, |j_\ell(k \eta)|^2$}

We rewrite the integral under consideration
\begin{equation}\label{approx:step1}
I_1(\ell, \eta) = \int_0^\infty \, dk \, k^2 \,  w(k) \, |j_\ell(k \eta)|^2 \, .   
\end{equation}
The first step consists in the change of the integration variable in Eq. \eqref{approx:step1}, defining the new variable $x$ through $k = x (\ell + 1/2)/\eta$. Integral \eqref{approx:step1} becomes
\begin{equation}\label{approx:step2}
I_1(\ell, \eta) = \frac{(\ell+1/2)^3}{\eta^3} \int_0^\infty \, dx \, x^2 \,  w\left(\frac{(\ell+1/2)}{\eta} x\right) \, |j_\ell\left((\ell+1/2)x\right)|^2 \, .    
\end{equation}
Now we consider the following Nicholson approximation of the spherical Bessel function which is valid for positive $x$ arguments and high-$\ell$ index (see e.g. \cite{NIST:DLMF})
\begin{equation} \label{j_approx}
j_\ell((\ell+1/2)x) \simeq \sqrt \pi \left(\frac{\xi(x)}{1-x^2}\right)^{1/4} \frac{\mbox{Ai}((\ell+1/2)^{2/3} \xi(x))}{(\ell+1/2)^{5/6} x^{1/2}} \, ,  
\end{equation}
where $\mbox{Ai}(z)$ is the Airy function and 
\begin{equation} \label{xi_def}
\xi(x) =
\begin{cases}
- \left(\frac{3}{2}\right)^{2/3} \left(\sqrt{x^2 - 1} - \arctan{\sqrt{x^2 -1}} \right)^{2/3}, & \text{if $x\geq 1$ \, ,} \\
\left(\frac{3}{2}\right)^{2/3} \left( \mbox{arctanh}{\sqrt{1 - x^2}} - \sqrt{1 - x^2}\right)^{2/3}, & \text{if $x\leq 1$ \, .}
\end{cases}
\end{equation}
Doing the modulus square of Eq. \eqref{j_approx}, we obtain 
\begin{equation} \label{j_approx2}
|j_\ell((\ell+1/2)x)|^2 \simeq \pi \frac{|\xi(x)|^{1/2}}{|x^2-1|^{1/2}} \frac{\mbox{Ai}^2((\ell+1/2)^{2/3} \xi(x))}{(\ell+1/2)^{5/3} x} \, .
\end{equation}
Moreover, we can expand the Airy function squared in \eqref{j_approx2} using the following integral representation (see e.g. \cite{vallee2004airy})
\begin{equation}\label{airy_integral}
\mbox{Ai}^2(z) = \frac{1}{2 \pi^{3/2}} \int_0^\infty \,\frac{dt}{\sqrt t} \, \cos\left(\frac{1}{12} t^3 + z t + \frac{\pi}{4}\right) \, ,    
\end{equation}
valid for a real variable $z$. Since $\ell$ is large in the approximated Eq. \eqref{j_approx2}, then the argument $z$ of the Airy function squared $\mbox{Ai}^2(z)$ is large. In this limit the $t^3$ term in the cosine of Eq. \eqref{airy_integral} can be dropped by the virtue of the Riemann-Lebesgue lemma, and we remain with
\begin{equation}\label{airy_integral2}
\mbox{Ai}^2(z) \simeq \frac{1}{2 \pi^{3/2}} \int_0^\infty \,\frac{dt}{\sqrt t} \, \cos\left(z t + \frac{\pi}{4}\right) = \frac{1}{2 \pi} \frac{1}{(-z)^{1/2}} \theta(-z) \, , 
\end{equation}
where $\theta(z)$ denotes the Heaviside step function.

Under this approximation, because of the appearing of the Heaviside function we have that $|j_\ell((\ell+1/2)x)|^2$ vanishes in the interval $0<x<1$, where $\xi(x)$ is positive (see Eq. \eqref{xi_def}). For $x \geq 1$, substituting Eq. \eqref{airy_integral2} into \eqref{j_approx2}, we get
\begin{equation} \label{j_approx3_x}
|j_\ell((\ell+1/2)x)|^2 \simeq  \frac{1}{2 (\ell + 1/2)^2} \frac{1}{x \sqrt{x^2 - 1}} \, .
\end{equation}
Thus, substituting Eq. \eqref{j_approx3_x} into \eqref{approx:step2}, we obtain
\begin{equation}\label{approx:step3}
I_1(\ell, \eta) \simeq \frac{(\ell+1/2)}{2 \eta^3} \int_1^\infty \, dx \, \frac{x}{\sqrt{x^2 - 1}} \,  w\left(\frac{(\ell+1/2)}{\eta} x\right) \, .    
\end{equation}
Finally, we introduce another change of variable, defining $x = \sqrt{1+y}$. So, integral \eqref{approx:step3} becomes
\begin{equation} \label{approx:step4}
I_1(\ell, \eta)|_{\ell \rightarrow \infty} \simeq  \frac{(\ell+1/2)}{4 \eta^3} \int_0^\infty \, \frac{dy}{\sqrt y} \,  w\left(\frac{(\ell+1/2)}{\eta} \sqrt{1+y}\right) \, .       
\end{equation}

\subsection{High-$\ell$ approximation of $\int_0^\infty \, dk \, k^2 \,  w(k) \, |j'_\ell(k \eta)|^2$}

The integral we want to approximate is
\begin{equation}\label{approx2:step1}
I_2(\ell, \eta) = \int_0^\infty \, dk \, k^2 \,  w(k) \, |j'_\ell(k \eta)|^2 \, .
\end{equation}
As we have done before, we switch to the variable $x$ defined through $k = x (\ell + 1/2)/\eta$, obtaining
\begin{equation}\label{approx2:step2}
I_2(\ell, \eta) = \frac{(\ell+1/2)^3}{\eta^3} \int_0^\infty \, dx \, x^2 \,  w\left(\frac{(\ell+1/2)}{\eta} x\right) \, |j'_\ell\left((\ell+1/2)x\right)|^2 \, .    
\end{equation}
Now, in order to find the high $\ell$ value of $j'_{\ell}(z)$, we need to differentiate Eq. \eqref{j_approx}. Thus, using the fact that by definition
\begin{equation}
\xi'(x) = - \frac{1}{x} \left(\frac{\xi(x)}{1-x^2}\right)^{-1/2}  \, ,  
\end{equation}
we find
\begin{equation} \label{j2_approx}
j'_\ell((\ell+1/2)x) \simeq - \sqrt \pi \left(\frac{\xi(x)}{1-x^2}\right)^{-1/4} \frac{\mbox{Ai}'((\ell+1/2)^{2/3} \xi(x))}{(\ell+1/2)^{7/6} x^{3/2}} \, .  
\end{equation}
In particular, we are interested to the modulus square of \eqref{j2_approx}, i.e.
\begin{equation} \label{j2_approx2}
|j'_\ell((\ell+1/2)x)|^2 \simeq  \pi \left(\frac{\xi(x)}{x^2-1}\right)^{-1/2} \frac{(\mbox{Ai}'((\ell+1/2)^{2/3} \xi(x)))^2}{(\ell+1/2)^{7/3} x^{3}} \, .  
\end{equation}
Using the Airy's equation $\mbox{Ai}''(z) = z \mbox{Ai}(z)$ (see e.g. \cite{vallee2004airy}), we can express the derivative of the Airy function squared as
\begin{equation}
(\mbox{Ai}'(z))^2 = - z \mbox{Ai}^2(z) + \frac{1}{2} \frac{d^2}{dz^2} \mbox{Ai}^2(z) \, , \end{equation}
which in turn, using the large z approximation of $\mbox{Ai}^2(z)$ \eqref{airy_integral2}, gives
\begin{equation} \label{Airy_derivative}
(\mbox{Ai}'(z))^2 \simeq \frac{1}{2 \pi} (- z)^{1/2} \theta(-z) \, .      
\end{equation}
As before, we have that $|j'_\ell((\ell+1/2)x)|^2$ vanishes in the interval $0<x<1$ where $\xi(x)$ is positive. For $x \geq 1$, substituting Eq. \eqref{Airy_derivative} into \eqref{j2_approx2}, we get
\begin{equation} \label{j2_approx3_x}
|j'_\ell((\ell+1/2)x)|^2 \simeq  \frac{1}{2 (\ell + 1/2)^2} \frac{\sqrt{x^2 - 1}}{x^3} \, .
\end{equation}
Thus, substituting Eq. \eqref{j2_approx3_x} into \eqref{approx2:step2}, we obtain
\begin{equation}\label{approx2:step3}
I_2(\ell, \eta) \simeq \frac{(\ell+1/2)}{2 \eta^3} \int_1^\infty \, dx \, \frac{\sqrt{x^2 - 1}}{x} \,  w\left(\frac{(\ell+1/2)}{\eta} x\right) \, .    
\end{equation}
Finally, we introduce again the change of variable $x = \sqrt{1+y}$, and we get
\begin{equation}
I_2(\ell, \eta)|_{\ell \rightarrow \infty} \simeq  \frac{(\ell+1/2)}{4 \eta^3} \int_0^\infty \, dy \, \frac{\sqrt y}{1+y} \,  w\left(\frac{(\ell+1/2)}{\eta} \sqrt{1+y}\right) \, .       
\end{equation}

\subsection{High-$\ell$ approximation of $\int_0^\infty \, dk \, k^2 \,  w(k) \, j_\ell(k \eta) \, j'_\ell(k \eta)$}

The last integral we want to approximate in the high-$\ell$ limit is the following
\begin{equation}\label{approx3:step1}
I_3(\ell, \eta) = \int_0^\infty \, dk \, k^2 \,  w(k) \, j_\ell(k \eta) \, j'_\ell(k \eta)  \, .
\end{equation}
We can rewrite this integral in terms of the derivative with respect to $\eta$ as
\begin{equation}\label{approx3:step2}
I_3(\ell, \eta) = \frac{1}{2} \frac{d}{d \eta} \int_0^\infty \, dk \, k^2 \,  \tilde w(k) \, j_\ell(k \eta) \, j_\ell(k \eta)  = \frac{1}{2} \frac{d}{d \eta} \tilde I_1(\ell, \eta) \, ,
\end{equation}
where we redefined the kernel function as $\tilde w(k) =  w(k)/k$, and $\tilde I_1(\ell, \eta)$ is integral \eqref{approx:step1} with $w(k)$ replaced by $\tilde w(k)$.
Now, it is enough to insert the approximated integral \eqref{approx:step4} into \eqref{approx3:step2} to get
\begin{align}\label{approx3:step3}
I_3(\ell, \eta)|_{\ell \rightarrow \infty} \simeq  &- \frac{1}{4 \eta^3} \int_0^\infty \,  \frac{dy}{\sqrt y \sqrt{1+y}} \,  w\left(\frac{(\ell+1/2)}{\eta} \sqrt{1+y}\right) \\
&- \frac{(\ell+1/2)}{8 \eta^4} \int_0^\infty \, \frac{dy }{\sqrt y} \,  w'\left(\frac{(\ell+1/2)}{\eta} \sqrt{1+y}\right) \, .
\end{align}

%\section{Alternative expression of the IA effect \label{second_align}}
%In Ref. \cite{Schmidt:2013gwa}, an alternative way to compute the contribution of tensor modes to galaxy alignments is considered. In fact, instead of considering the effects of alignment at the source redshift, the authors computed the higher (second) order contribution to the galaxy matter overdensity from tensor perturbations, assuming that the form of the final IA is the same as this second order contribution. In the following we quantitatively review the computations of this paper, assuming to be in a matter dominated universe.

\bibliographystyle{utphys}
\bibliography{references.bib}

\end{document}